\documentclass[11pt, a4paper]{amsart}
\usepackage[dvips]{epsfig}
\usepackage{amsmath}
\usepackage{amssymb}
\usepackage{amsfonts}
\usepackage{amsthm}
\usepackage{amsbsy}
\usepackage{amsgen}
\usepackage{amscd}
\usepackage{amsopn}
\usepackage{amstext}
\usepackage{amsxtra}

\newtheorem{theorem}{Theorem}[section]
\newtheorem{proposition}[theorem]{Proposition}
\newtheorem{lemma}[theorem]{Lemma}
\newtheorem{corollary}[theorem]{Corollary}

\theoremstyle{definition}
\newtheorem{definition}[theorem]{Definition}
\newtheorem{notation}[theorem]{Notation}
\newtheorem{remark}[theorem]{Remark}

\newcommand {\Z} {\mathbb{Z}}

\newcommand {\s} {\mathcal{S}}
\newcommand {\R} {\mathbb{R}}

\newcommand {\E} {\mathbb{E}}

\newcommand {\PP} {\mathcal P}

\newcommand {\eigspc} {\mathcal{E}_n}
\newcommand {\eigspcdim} {\mathcal{N}}
\newcommand {\eigval} {E_{n}}
\newcommand {\spheredim}{m}
\newcommand {\sphere}{\s^\spheredim}

\newcommand {\funcvalcorrinv} {\frac{|\sphere|}{\eigspcdim}}
\newcommand {\funcdercorr} {\frac{\eigval}{\spheredim}}
\newcommand {\funcdercorrinv} {\frac{\spheredim}{\eigval}}
\newcommand {\ultsphpol} {Q_{n}^{\spheredim}}

\newcommand{\leray}{\mathcal {L}}
\newcommand{\length}{\mathcal{Z}}
\newcommand{\var}{\operatorname{Var}}

\newcommand{\tr}{\operatorname{tr}}
\newcommand{\vol}{\operatorname{vol}}

\newcommand{\meas}{\operatorname{meas}}

\begin{document}

\title[Distribution of the nodal sets]
{On the distribution of the nodal sets of random spherical harmonics}
\author{Igor Wigman}
\address{Centre de recherches math\'ematiques (CRM),
Universit\'e de Montr\'eal C.P. 6128, succ. centre-ville Montr\'eal,
Qu\'ebec H3C 3J7, Canada} \email{wigman@crm.umontreal.ca}
\thanks{The author is supported by a CRM ISM fellowship}

\maketitle

\begin{abstract}
We study the volume of the nodal set of eigenfunctions of the
Laplacian on the $\spheredim$-dimensional sphere. It is well known
that the eigenspaces corresponding to $\eigval=n(n+\spheredim-1)$
are the spaces $\eigspc$ of spherical harmonics of degree $n$, of
dimension $\eigspcdim$. We use the multiplicity of the eigenvalues
to endow $\eigspc$ with the Gaussian probability measure and study
the distribution of the $\spheredim$-dimensional volume of the nodal
sets of a randomly chosen function. The expected volume is
proportional to $\sqrt{\eigval}$. One of our main results is
bounding the variance of the volume to be
$O(\frac{\eigval}{\sqrt{\eigspcdim}})$.

In addition to the volume of the nodal set, we study its Leray
measure. We find that its expected value is $n$ independent. We are
able to determine that the asymptotic form of the variance is
$\frac{const}{\eigspcdim}$.
\end{abstract}

\tableofcontents

\section{Introduction}

Let $M$ be a smooth compact manifold and $f$ a real valued function
on $M$. We define its nodal set to be the subset of $M$, where $f$
vanishes and we are interested mainly in the nodal sets of
eigenfunctions of the Laplacian on $M$. It is known ~\cite{Cheng},
that generically, the nodal sets are smooth submanifolds of $M$ with
codimension $1$. For example, if $M$ is a surface, the nodal sets
are {\em lines}. One is interested in studying their volume (i.e.
the length of the nodal line for the $2$-dimensional case), and
other nodal properties for highly excited eigenstates. It was
conjectured by Yau that the volume of the nodal set is bounded from
above and below by a multiple of the square root of the Laplace
eigenvalue. The lower bound was proven by Bruning and Gromes
~\cite{Bruning-Gromes} and Bruning ~\cite{Bruning} for the planar
case. Donnelly-Fefferman's celebrated result
~\cite{Donnelly-Fefferman} resolved Yau's conjecture for real
analytic metrics. However the general case of a smooth manifold is
still open.

\subsection{Spherical Harmonics}

In this paper, we study the nodal sets for the eigenfunctions of the
Laplacian $\Delta$ on the $\spheredim$-dimensional unit sphere
$\sphere$. It is well known that the eigenvalues $E$ of the Laplace
equation
\begin{equation*}
\Delta f +E f = 0
\end{equation*}
on $\sphere$ are all the numbers of the form
\begin{equation}
\label{eq:eigval def} E = \eigval = n(n+\spheredim-1),
\end{equation}
where $n$ is an integer. Given a number $\eigval$, the corresponding
eigenspace is the space $\eigspc$ of the spherical harmonics of
degree $n$. Its dimension is given by
\begin{equation}
\label{eq:eigspcdim asymp} \eigspcdim = \eigspcdim_{n}=
\frac{2n+\spheredim-1}{n+\spheredim-1} {n+\spheredim-1 \choose
\spheredim -1} \sim \frac{2}{(\spheredim-1)!}n^{\spheredim-1}.
\end{equation}
Given an integral number $n$, we fix an $L^2(\sphere)$ orthonormal
basis of $\eigspc$
\begin{equation*}
\eta_{1}(x) = \eta_{1}^{n} (x), \, \eta_{2}(x) = \eta_{2}^{n}
(x),\ldots ,\eta_{\eigspcdim}(x) = \eta_{\eigspcdim}^{n} (x),
\end{equation*}
giving an identification $\eigspc\cong\R^{\eigspcdim}$. For further reading on the spherical harmonics we refer the reader
to ~\cite{AAR}, chapter 9.

\subsection{Random model}

We consider a {\em random eigenfunction}
\begin{equation}
\label{eq:rand eigfnc def} f(x)= \sqrt{\funcvalcorrinv}
\sum\limits_{k=1}^{\eigspcdim} a_{k}\eta_{k}(x),
\end{equation}
where $a_k$ are Gaussian $N(0,1)$ i.i.d. which we assume to be
defined on the same sample space $\Omega$. That is, we use the
identification $\eigspc\cong\R^{\eigspcdim}$ to induce the Gaussian
probability measure $\upsilon$ on $\eigspc$ as
\begin{equation*}
d\upsilon(f) = e^{-\frac{1}{2}\|\vec{a}\|^2}\frac{da_{1} \cdot\ldots\cdot
da_{\eigspcdim}}{(2\pi)^{\eigspcdim/2}},
\end{equation*}
where $\vec{a} = (a_{i})\in\R^{\eigspcdim}$ are as in \eqref{eq:rand
eigfnc def}.

Note that $\upsilon$ is invariant with respect to the
orthonormal basis for $\eigspc$. As usual, for any random variable
$X$ on $\Omega$, we denote its expectation $\E X$. For example, with
the normalization factor in \eqref{eq:rand eigfnc def}, for every
{\em fixed} point $x\in\sphere$, one has
\begin{equation}
\label{eq:E(f(x)^2)=1} \E [f(x)^2] =
\funcvalcorrinv\sum\limits_{i=1}^{\eigspcdim} \eta_i(x) ^2 = 1,
\end{equation}
a simple corollary from the addition theorem (see
\eqref{eq:E(f(x)^2)=1 pr}).

Any characteristic $X(L)$ of the nodal set
$$L=L_{f}=\{x\in\sphere:\: f(x)=0 \} $$ is a random variable
defined on the same sample space $\Omega$. We are interested in the
distribution of two different characteristics. The most natural
characteristic of the nodal set $L_{f}$ of $f$ is, of course, its
$(\spheredim-1)$-dimensional volume $\length=\length(f)$. The study of
the distribution
of the random variable $\length$ for a random $f\in\eigspc$ is one of the goals
of the present paper.

Berard ~\cite{Berard} showed that the expected volume $\E\length$ is
\begin{equation*}
\E \length(f) = const\cdot \sqrt{\eigval}
\end{equation*}
(see proposition \ref{prop:exp len}) and
Neuheisel ~\cite{Neuheisel} proved that as $n\rightarrow\infty$,
\begin{equation}
\label{eq:var bnd Neuheisel}
\var(\length) = O\bigg(\frac{\eigval}{n^{\frac{(\spheredim-1)^2}{3\spheredim+1}}}\bigg) =
O\bigg(\frac{\eigval}{\eigspcdim^{\frac{\spheredim-1}{3\spheredim+1}}}\bigg).
\end{equation}

\begin{remark}
\label{rem:Neuh sphr mod}
Rather than
taking $a_{k}$ standard Gaussian i.i.d., Neuheisel assumes that the vector $\vec{a}=(a_{k})\in\R^{\eigspcdim}$
is chosen uniformly on the unit sphere $\s^{\eigspcdim-1}$. However, it is easy to see,
that, since $\length(f)=\length(b\cdot f)$ for any constant $b\in\R$, both of those models
are equivalent.
\end{remark}

The volume of the nodal line of a random eigenfunction on the torus
$$\mathcal{T}^{\spheredim} = \R^{\spheredim}/\Z^{\spheredim}$$ was
studied by Rudnick and Wigman ~\cite{RW}. In this case, it is not
difficult to see that the expectation is given by $\E \length
(f^{\mathcal{T}^\spheredim}) = const\cdot \sqrt{E}$. Moreover, they
prove that as the eigenspace dimension $\eigspcdim$ grows to
infinity, the variance is bounded by $$\var
\length(f^{\mathcal{T}^\spheredim})  =
O\bigg(\frac{E}{\sqrt{\eigspcdim}}\bigg),$$ which, in particular,
implies that the tails of the distribution of the normalized random
variable $\frac{\length}{\E \length}$ die.

More generally, one may also consider a random model of eigenfunctions for a generic compact manifold $M$.
Of course, for generic manifolds, one does not expect the Laplacian to have any multiplicities, so that
we cannot introduce a Gaussian ensemble on the eigenspace.
Let $E_{j}$ be the eigenvalues and $\phi_{j}$ the corresponding eigenfunctions. It is well known that
the $E_{j}$ are discrete, $E_{j}\rightarrow\infty$ and $L^{2}(M) = span\{\phi_{j} \}$.

In this case, rather than considering random eigenfunctions, one considers random
{\em combinations} of eigenfunctions with growing energy window of either type
\begin{equation*}
f^{L}(x) = \sum\limits_{E_{j}\in [0,E]} a_{j}\phi_{j}(x)
\end{equation*}
(called the long range), or
\begin{equation*}
f^{S}(x) = \sum\limits_{\sqrt{E_{j}}\in [\sqrt{E},\sqrt{E}+1]} a_{j}\phi_{j}(x),
\end{equation*}
(called the short range), as $E\rightarrow\infty$.
Berard ~\cite{Berard} found that $$\E\length(f^{L}) \sim c_{M} \cdot \sqrt{E}$$ and recently Zelditch
~\cite{Z1} proved that $$\E\length(f^{S}) \sim c_{M} \cdot \sqrt{E},$$ notably with the same constant
$c_{M}$ for both the long and the short ranges.

Berry \cite{Berry 2002} computed the expected length of nodal lines
for isotropic, monochromatic random waves in the plane, which are
eigenfunctions of the Laplacian with eigenvalue $\eigval$. He found
that the expected length (per unit area) is again of size
approximately $\sqrt{\eigval}$ and argued that the variance should
be of order $\log \eigval$.

\vspace{3mm}

\subsection{Leray nodal measure}
\label{sec:ler meas def}

Another property of the nodal line we consider
is its {\em Leray measure} (also called {\em the microcanonical measure}).
Given a function $f$ on $\sphere$, we define the Leray nodal measure
to be
\begin{equation}\label{eq:leray meas def}
\leray(f):=\lim\limits_{\epsilon\to 0} \frac 1{2\epsilon} \meas
\{x\in \sphere: |f(x)|<\epsilon \},
\end{equation}
provided that the last limit exists. One may write the definition
\eqref{eq:leray meas def} of the Leray nodal measure formally as
\begin{equation}\label{eq:leray meas form}
\leray(f):=\int\limits_{\sphere} \delta(f(x)) dx,
\end{equation}
where $\delta$ is the Dirac delta function.

As is well known, the limit \eqref{eq:leray meas def} exists when
$\nabla f\neq 0$ on the nodal set in which case
$$\leray(f) = \int_{\{x:f(x)=0\}} \frac{d\nu'(x)}{|\nabla f(x)|},
$$
where $\nu'$ is the Riemannian hypersurface measure on the nodal
set. This holds almost always on $\eigspc$ (see section
\ref{sec:sing func}).

The distribution of the Leray nodal measure on the sphere was also
considered by Neuheisel. As in case of the volume, one may compute
the expected value
\begin{equation*}
\E \leray = \frac{|\sphere |}{\sqrt{2\pi}}
\end{equation*}
using a rather standard computation (see proposition \ref{prop:exp ler meas}) and
Neuheisel proved that the variance is bounded by
\begin{equation}
\label{eq:var Ler bnd Neuheisel}
\var\leray = O\bigg(\frac{1}{n^{\frac{\spheredim-1}{2}}}\bigg) = O(\frac{1}{\sqrt{\eigspcdim}})
\end{equation}

\begin{remark}
Here, as in the case for the volume, Neuheisel considered a slightly
different variation of the random model (see remark \ref{rem:Neuh
sphr mod}). Even though the Leray nodal measure is not invariant
under dilations, i.e. $$\leray(b\cdot f) = \frac{1}{b}\leray(f),$$
those models are still equivalent {\em asymptotically}, as
$\eigspcdim\rightarrow\infty$.
\end{remark}

The Leray measure $\leray(f^{\mathcal{T}^\spheredim})$ for the
random eigenfunctions on the torus $\mathcal{T}^{n}$ was considered
by Oravecz, Rudnick and Wigman ~\cite{ORW}. The expectation is given
by
\begin{equation*}
\E \leray(f^{\mathcal{T}^\spheredim}) = \frac{1}{\sqrt{2\pi}}.
\end{equation*}
These authors
were able to establish the variance to be asymptotic to
\begin{equation*}
\var\leray(f^{\mathcal{T}^\spheredim})\sim c\cdot \frac{1}{\eigspcdim}
\end{equation*}
for some $c>0$, for $m=2$ and $m \ge 5$.

\subsection{The expectation}

\begin{proposition}
\label{prop:exp ler meas} For $n$ sufficiently large, the expectation of
the Leray nodal measure of the random eigenfunction is given by
\begin{equation*}
\E\leray (f) = \frac{|\sphere|}{\sqrt{2\pi}}.
\end{equation*}
\end{proposition}

\begin{proposition}
One has
\label{prop:exp len}
\begin{equation}
\label{eq:Elen=c sqrt(E)} \E\length(f) = c_{\spheredim}
\cdot\sqrt{\eigval},
\end{equation}
with the constant $c_{\spheredim}$ defined by
\begin{equation}
\label{eq:c def} c_{\spheredim} =
\frac{2\pi^{\spheredim/2}}{\sqrt{\spheredim}\Gamma(\frac{m}{2})}.
\end{equation}
\end{proposition}

\subsection{Statement of the main results}
Our main results concern the variance of the Leray nodal measure
$\leray$ and the volume $\length$ of the nodal set. We improve on
Neuheisel's results \eqref{eq:var bnd Neuheisel} and \eqref{eq:var
Ler bnd Neuheisel}, and need to use some of the steps in his work;
however because some of the arguments in Neuheisel contain gaps, we
need to redo them, partially accounting for the length of this
paper.

\vspace{2mm}

\noindent For $\leray$ we were able to determine its asymptotics precisely.
\begin{theorem}
\label{thm:var ler meas} As $n\rightarrow\infty$, the variance of
the Leray nodal measure is asymptotic to
\begin{equation}
\label{eq:var ler meas} \var\leray(f) \sim
\frac{2^{\spheredim-2}\pi^{\frac{\spheredim-2}{2}}\Gamma(\frac{\spheredim}{2})
|\sphere |}{(\spheredim-1)!} \cdot \frac{1}{\eigspcdim}.
\end{equation}
\end{theorem}
\noindent One should compare the asymptotic result \eqref{eq:var ler meas} to Neuheisel's bound
\eqref{eq:var Ler bnd Neuheisel}.

\begin{remark}
Note that unlike the torus, our proof here works for any dimension
$\spheredim \ge 2$, including $\spheredim =3,\, 4$. The reason is
that for the sphere, the so-called two point function $u$ (to be
defined, see \eqref{eq:u(x) def}) is related to the ultraspherical
polynomials, a standard family of orthogonal polynomials ~\cite{SZ}.
In particular, using Hilb's asymptotics for the ultraspherical
polynomials, it is easy to show that the 4th moment of $u$ is
dominated by its second moment (see lemmas ~\ref{lem:2nd mom Qn} and
~\ref{lem:4th mom Qn}).

Unlike the spherical case, the two point function for the
$d$-dimensional torus is related to the distribution of points
$$\vec{n} = (n_{1},\ldots n_{d})\in \Z^d$$ so that $$\| \vec{n} \|^2
= n_{1}^2+\ldots +n_{d}^2 = \frac{E}{4\pi^2}.$$ For $d \ge 5$ a
strong equidistribution result for $\vec{n}$ implies in particular
that the 4th moment of $u$ is dominated by its second moment. For
the two-dimensional case we used a special result due to Zygmund.
The remaining cases $d=3,\, 4$ are, to our best knowledge, open.
\end{remark}

\vspace{3mm} Concerning the volume, we have the following result:
\begin{theorem}
\label{thm:var length} One has
\begin{equation}
\label{eq:var length} \var\length(f) =
O\bigg(\frac{\eigval}{\sqrt{\eigspcdim}}\bigg),
\end{equation}
asymptotically as $n\rightarrow\infty$.
\end{theorem}

Note that theorem \ref{thm:var length} implies that the variance of
the {\em normalized} random variable
$\tilde{\length}:=\frac{\length}{\E\length}$ with expectation $1$,
vanishes as $n\rightarrow\infty$. Thus, in particular, the tails of
the distribution of $\tilde{\length}$ ``die", that is, for every
$\epsilon>0$, most of the mass of $\tilde{\length}$ is concentrated
in $[1-\epsilon, 1+\epsilon]$. In addition, theorem \ref{thm:var
length} bounds the ``typical" size of the tail of the distribution
of $\tilde{\length}$ (and thus of $\length$). One should compare
\eqref{eq:var length} to \eqref{eq:var bnd Neuheisel}, obtained by
Neuheisel. Partially motivated by the recent result ~\cite{GW} for
the analogous ensemble of random one dimensional trigonometric
polynomials, it may be possible to improve the bound to
$\eigval/\eigspcdim$ .

\subsection{On the proofs of the main results}

The spherical case offers some marked differences from that of the torus ~\cite{ORW} and
~\cite{RW}. Unlike the torus, which is identified with the unit square with its sides
pairwise glued, the sphere possesses a nontrivial geometry. In the course of the proof
of the main results, one has to study the joint distribution of the gradients $\nabla f(x)$
and $\nabla f(y)$ as random vectors, where $x,y\in \sphere$ are {\em fixed}, and $f\in\eigspc$ is randomly
chosen. The main obstacle here is that for different points $x\in\sphere$, the gradients
live in different spaces, namely, the tangent spaces $T_{x}(\sphere)$ which are, in general, different.

One then has to canonically identify the spaces $T_{x}(\sphere)$ via a family of isometries $\phi_{x}$ smooth
w.r.t. $x$. In reality, such a choice is not possible for every $x$,
and we treat this complication in section \ref{sec:orthonorm bas corr mat exp}.

Once the geometric problems are resolved, the treatment of the so-called {\em two-point} (or,
alternatively, {\em covariance}) function and its derivatives is more standard, related to the well-known ultraspherical polynomials
(see ~\cite{SZ} or appendix \ref{sec:ultrasph pol}).
In particular, we find that the geometrical structure
of the so-called {\em singular sets} on $\sphere$ is less complicated than the singular set on the torus
(see section \ref{sec:sing set}).

\subsection{Some Conventions}
Throughout the paper, the letters $x,y$ and $z$ will denote either
points on the sphere $\sphere$ or spherical variables and $t$ will
denote a real variable. For $x,\, y\in\sphere$, $d(x,y)$ will stand
for the spherical distance between $x$ and $y$. The letters $\mu$,
$\nu$, $\upsilon$ will be reserved for measures, where the measure
$\nu$ will stand for the uniform measure on $\sphere$ so that
$d\nu(x) = dx$.

Finally, given a set $S$, we denote its volume by $|S|$. For example,
\begin{equation}
\label{eq:sphere vol}
|\sphere| = \frac{2\pi^{\frac{\spheredim+1}{2}}}{\Gamma(\frac{\spheredim+1}{2})}.
\end{equation}
In this manuscript we will use the notations $A\ll B$ and $A=O(B)$
interchangeably.

\subsection{Plan of the paper}
This paper is organized as follows. Section \ref{sec:expectation} is
devoted to the computation of the expected value of the Leray nodal
measure and the volume, that is, proving propositions \ref{prop:exp
ler meas} and \ref{prop:exp len}, where the rest of the paper
focuses on the variance of those characteristics, i.e. proving
theorems \ref{thm:var ler meas} and \ref{thm:var length}. The
treatment of the variance in both cases will be divided into two
steps. First, we express it in an integral form in section
\ref{sec:int form sec mom}. We treat the integrals obtained in
section \ref{sec:int form sec mom} throughout section \ref{sec:asymp
var}. In case of the Leray measure, we will be able to give a
precise asymptotic expression. In the case of volume, we give an
upper bound.

Appendix \ref{sec:ultrasph pol} will introduce the reader to the
ultraspherical polynomials and will also provide all the necessary
background we will need in this paper. The goal of appendix
\ref{sec:sing func rare} is to prove that the set of ``bad"
(singular) eigenfunctions in the space of all the eigenfunctions, is
``rare" in some strong sense. Finally, appendix \ref{sec:f (x)(y)gr
f(x)(y) sp} will prove a particular nondegeneracy result for the
distribution of the eigenfunctions and its gradients, needed to give
meaning to the integral formula obtained for the variance of the
volume given in section \ref{sec:int form sec mom}.

\subsection{Acknowledgements}
The author wishes to thank Ze\'{e}v Rudnick for initiating this
research and for his help and support while conducting it. Many
stimulating discussions with Mikhail Sodin, Dmitry Jakobson and
St\'{e}phane Nonnenmacher are appreciated. The author is grateful to
Sherwin Maslowe for proofreading this paper. I would like to thank
CRM Analysis laboratory and its members for their support. Some part
of this research was done during the author's visit to the Bielefeld
University, supported by SFB 701: Spectral Structures and
Topological Methods in Mathematics. Finally, I wish to thank the
anonymous referee for his comments and suggestions.

\section{Expectation}
\label{sec:expectation} In this section we prove propositions
\ref{prop:exp ler meas} and \ref{prop:exp len}. As a start, we wish
to stay away from the set of the singular functions discussed in
section \ref{sec:sing func}.

\subsection{The singular functions}
\label{sec:sing func}

In this section we define the notion of the singular functions and
formulate the intuitive statement that they are ``rare". The proofs
are given in appendix \ref{sec:sing func rare}.

\begin{definition}
\label{def:sing func} An eigenfunction $f\in \eigspc$ is {\em
singular} if $\exists x\in \sphere$ with  $f(x)=0$  and $\nabla f(x)
= \vec{0}$. An eigenfunction $f\in \eigspc$ is  {\em nonsingular} if
$\nabla f\neq \vec 0$ on the nodal set.
\end{definition}

A nonsingular eigenfunction has no self-intersections. We denote
$Sing\subseteq \eigspc$ to be the set of all the singular
eigenfunctions. First, we claim that as a set, $Sing$ is {\em
``small"}.

\begin{lemma}
\label{lem:Sing codim 1} The set $Sing$ has codimension $1$ in
$\eigspc$.
\end{lemma}

Now, given $x\in\sphere$ and $b\in\R$, we denote $\PP_{b}^{x}$ to be
the set of all the eigenfunctions which attain the value $b$ at the
point $x$. That is,
\begin{equation}
\label{eq:PPxa def} \PP^{x}_{b} = \{f\in\eigspc:\: f(x) = b \}.
\end{equation}
The set $\PP^{x}_{b}$ is a hyperplane in $\eigspc$.

Moreover, given $(x,y)\in{\sphere}\times\sphere$ and
$b=(b_1,b_2)\in\R^2$ we denote
\begin{equation}
\label{eq:PPxya def} \PP^{x,y}_{b} = \{f\in\eigspc:\: f(x) =
b_{1},\, f(y) = b_{2} \}.
\end{equation}
For $x\ne \pm y$, $\PP^{x,y}_{b}$ is an affine subspace of $\eigspc$
of codimension $2$, as it is easy to see from the addition theorem
(see section \ref{eq:two-pnt func}).

The next couple of lemmas establish the fact that the intersections
of $Sing$ with $\PP^{x}_{b}$ and $\PP^{x,y}_{b}$ for $x\ne \pm y$,
are of codimension $1$. Lemma \ref{lem:Sing codim 1 Paxy} is
essential while treating the variance of the Leray nodal measure
(section \ref{sec:ler var int form}).

\begin{lemma}
\label{lem:Sing codim 1 Pax} For every $x\in\sphere$ and $b\in\R$,
the set $$Sing_{b}^x := Sing\cap \PP_{b}^{x}$$ has codimension $1$
in $\PP_{b}^{x}$.
\end{lemma}

\begin{lemma}
\label{lem:Sing codim 1 Paxy} If $x,y\in\sphere$ and $x\ne \pm y$,
then for every $b=(b_1,b_2)\in\R^2$, the set $Sing_{b}^{x,y} :=
Sing\cap \PP_{b}^{x,y}$ has codimension $1$ in $\PP_{b}^{x,y}$.
\end{lemma}
The proofs of all the lemmas of this section are given in
appendix \ref{sec:sing func rare}.

\subsection{Two-point function}
\label{eq:two-pnt func} We define the so called {\em two-point}
function, also referred in the literature as the {\em covariance}
function
\begin{equation}
\label{eq:u(x,y) def} u(x,y) =u^{m}_{n}(x,y)=\E \big[ f(x)f(y)\big]
= \frac{|\sphere|}{\eigspcdim}\sum\limits_{k=1}^{\eigspcdim}\eta_{k}
(x) \eta_{k} (y).
\end{equation}

The addition theorem ~\cite{AAR}, page 456, theorem 9.6.3 implies
that
\begin{equation}
\label{eq:u(x,y) def ult} u(x,y) = \ultsphpol (\cos{d(x,y)}),
\end{equation}
where $$\ultsphpol :[-1,1]\rightarrow\R$$ are the {\em normalized}
ultraspherical polynomials defined and studied in appendix
\ref{sec:ultrasph pol}. Recall that $d(x,y)$ is the spherical distance so
that $$\cos{d(x,y)} = \langle x,y\rangle,$$ thinking of $\sphere$ as
being embedded into $\R^{\spheredim+1}$.

It is immediate that $u$ is rotationally invariant, i.e.
\begin{equation}
\label{eq:rot inv u} u(Rx,Ry) = u(x,y),
\end{equation}
where $R$ is any rotation on $\sphere$. In case $y$ is not
specified, it is taken to be the northern pole $N\in\sphere$, that
is
\begin{equation}
\label{eq:u(x) def} u(x) := u(x,N).
\end{equation}

For every $t\in [-1,1]$, $|\ultsphpol (t) | \le 1$ and
$|\ultsphpol(t)| =1$, if and only if $t=\pm 1$. Therefore
\begin{equation}
\label{eq:u(x,y)=1 iff x=pm y} (u(x,y)=\pm 1) \Leftrightarrow (x=\pm
y),
\end{equation}
and
\begin{equation}
\label{eq:u(x)=1 iff x=N,S} (u(x)= \pm 1 ) \Leftrightarrow (x\in
\{N, S\} ),
\end{equation}
where $N$ and $S$ are the northern and the southern poles
respectively.

\subsection{Leray nodal measure}

We will need the following definitions from ~\cite{ORW}, section 3.

For $\epsilon>0$, set
$$
\leray_\epsilon(f):=\frac 1{2\epsilon}\meas\{x: |f(x)|<\epsilon\}\;.
$$
so that $\leray(f) = \lim_{\epsilon\to 0} \leray_\epsilon(f)$.

For $\alpha>0$,  $\beta>0$ let
\begin{equation*}
\eigspc(\alpha,\, \beta) = \{f\in \eigspc :\: |f(x)| \leq \alpha
\Rightarrow\ |\nabla f (x)| > \beta \} \;.
\end{equation*}

The sets $\eigspc(\alpha,\, \beta)$ are open, and have the
monotonicity property
$$
\alpha_1 >\alpha_2 \Rightarrow \eigspc(\alpha_1,\beta) \subseteq
\eigspc(\alpha_2,\beta)
$$
 and
$$
\beta_1>\beta_2 \Rightarrow \eigspc(\alpha,\beta_1) \subseteq
\eigspc(\alpha,\beta_2) \;.
$$
Moreover, for any sequence $\alpha_n,\beta_n \to 0$ we have
$$\eigspc\setminus Sing =
\bigcup\limits_{n} \eigspc(\alpha_n,\, \beta_n) \;.
$$

We have (cf. ~\cite{ORW}, lemma 3.1)
\begin{lemma}\label{lem:ler meas bnd}
For $f\in \eigspc(\alpha,\, \beta)$ and $0<\epsilon < \alpha$, we
have
\begin{equation*}
\leray_\epsilon(f) \ll \sqrt{\eigval}
\end{equation*}
where the constant involved in the $'\ll'$-notation depends only on
$\alpha$ and $\beta$.
\end{lemma}

To prove lemma \ref{lem:ler meas bnd}, we will need lemma 3.2 from
~\cite{ORW}.
\begin{lemma}[Lemma 3.2 from ~\cite{ORW}]
\label{lem:Kac} Let $g(t)$ be a trigonometric polynomial on
$[0,2\pi]$ of degree at most $M$ so that there are $\alpha>0$,
$\beta>0$ such that $|g'(t)|>\beta$ whenever $|g(t)|<\alpha$. Then
for all $0<\epsilon<\alpha$ we have
$$
\frac 1 {2\epsilon}\meas\{t\in [0,2\pi]: |g(t)|<\epsilon \} \ll
\frac{M}{\beta},
$$
where the constant in the $'\ll'$-notation may depend on
$\spheredim$ only.
\end{lemma}

\begin{proof}[Proof of lemma \ref{lem:ler meas bnd}]
Let $(\phi_{1},\ldots \phi_{\spheredim})$ be the standard
multi-dimensional spherical coordinates so that $x\in\sphere$ is
parameterized by
\begin{equation*}
x = (\cos{\phi_{1}},\sin{\phi_{1}} \cos{\phi_{2}},\,\ldots,\,
\sin{\phi_{1}}\ldots \sin{\psi_{\spheredim}})
\end{equation*}
for $(\phi_{1},\ldots \phi_{\spheredim}) \in R:=[0,\pi]\times \ldots
[0,\pi]\times [0,2\pi]$. It is well-known that for $\phi_{i} \ne
0,\pi,2\pi$, $\big\{ \frac{\partial}{\partial \phi_{k}} \big\}$ is
an orthogonal basis of $T_{x}(\sphere)$ and we have
\begin{equation*}
\bigg\| \frac{\partial}{\partial \phi_{k}}\bigg\| = \sin{\phi_{1}}
\cdot \ldots\cdot \sin{\phi_{k-1}},
\end{equation*}
so that the Jacobian $$J=J(\phi_{1},\,\ldots,\,\phi_{\spheredim}
)=\frac{D x}{D(\phi_{1},\ldots,\, \phi_{\spheredim})}$$ satisfies
\begin{equation*}
J = \sin\phi_{1} ^{\spheredim-1}\cdot \sin{\phi_{2}}
^{\spheredim-2}\cdot\ldots\cdot \sin\phi_{\spheredim-1}.
\end{equation*}

Let $0<\epsilon <\alpha$. We write $$\meas \{x\in\sphere:\: |f(x)|
<\epsilon\}$$ as an integral
\begin{equation}
\label{eq:meas f<eps sph int}
\begin{split}
\meas \{x\in\sphere:\: |f(x)| <\epsilon\} = \int\limits_{\sphere}
\chi\bigg(\frac{f(x)}{\epsilon} \bigg) dx =
\int\limits_{A_{\epsilon} }
|J(\phi_{1},\,\ldots,\,\phi_{\spheredim})| d\phi_{1}\cdot\ldots\cdot
d\phi_{\spheredim}
\end{split}
\end{equation}
in the spherical coordinates, where we denoted $$A_{\epsilon} := \{
P \in R  :\: |f(P)| < \epsilon\}. $$ For $P\in R$, $1\le k \le
\spheredim$ we define $p_k(P) = \frac{1}{\big\|
\frac{\partial}{\partial \phi_{k}}\big\|} \frac{\partial f}{\partial
\phi_{k}} (P)$, so that
\begin{equation}
\label{eq:grad=p1^2+p2^2} \|\nabla f(x) \| ^2 = p_{1}^2 + p_{2}
^2+\ldots + p_{\spheredim}^2.
\end{equation}

We decompose $$A_{\epsilon} = W_{1} \cup W_{2}\cup\ldots \cup
W_{\spheredim}$$ with
$$W_{k} :=  \big\{P\in A_{\epsilon}:\: |p_{k} (P)| = \max\limits_{j} {|p_{j} (P)|}
\big\}.$$ Note that on $W_{k}$, $$|p_{i} (P)| >
\frac{\beta}{\sqrt{\spheredim}}, $$ by \eqref{eq:grad=p1^2+p2^2} and
$\|\nabla f(x)\| > \beta$ on $A_{\epsilon}$.

Note that for $\phi_{k}$, $k\ne k_{0}$ fixed,
$g(\phi_{k_{0}}):=f(\phi_{1},\,\ldots ,\, \phi_{\spheredim})$ is a
trigonometric polynomial in $\phi_{k_{0}}$ on either $[0,\pi]$ or
$[0,2\pi]$ of degree $\le n\le\sqrt{\eigval}$ with derivative
\begin{equation*}
g'(\phi) = \bigg\| \frac{\partial}{\partial
\phi_{k_{0}}}\bigg\|\cdot p_1 (P)
\end{equation*}
so that on $W_{k_{0}}$, $|g'(\phi)| > \big\|
\frac{\partial}{\partial
\phi_{k_{0}}}\big\|\frac{\beta}{\sqrt{\spheredim}}$. Thus lemma
\ref{lem:Kac} implies
\begin{equation*}
\meas \{\theta :\: |g(\phi_{k_{0}})| < \epsilon \} \ll
\frac{\sqrt{\eigval}}{\big\| \frac{\partial}{\partial
\phi_{k_{0}}}\big\|   }\cdot \epsilon.
\end{equation*}

Therefore the contribution of $W_{k_{0}}$ to the integral
\eqref{eq:meas f<eps sph int} is
\begin{equation*}
\begin{split}
&\int\limits_{W_{1} } |J| d{\phi_{1}}\cdot\ldots\cdot
d{\phi_{\spheredim}} \le \int\limits \meas \{\phi :\:
|g(\phi_{k_{0}})| < \epsilon \} d\phi_{1} \cdot \ldots
\hat{d\phi_{k_{0}}} \ldots d\phi_{\spheredim} \\&\ll \int\limits
\frac{|J|}{\big\| \frac{\partial}{\partial \phi_{k_{0}}}\big\|}
\cdot \sqrt{\eigval}\epsilon d\phi_{1} \cdot \ldots
\hat{d\phi_{k_{0}}} \ldots d\phi_{\spheredim} \ll \epsilon
\sqrt{\eigval}.
\end{split}
\end{equation*}
which concludes the proof of the lemma.

\end{proof}

We conclude the section with a formal derivation of
proposition \ref{prop:exp ler meas}. A rigorous proof proceeds along
the same lines as the proof of theorem 4.1 in ~\cite{ORW} (see
section 4.2), using lemmas \ref{lem:ler meas bnd}, \ref{lem:Sing
codim 1} and \ref{lem:Sing codim 1 Pax}. We omit it here.

\begin{proof}[Formal derivation of proposition \ref{prop:exp ler
meas}]

Given a function $f\in\eigspc$, we write its Leray nodal measure
formally as
\begin{equation*}
\leray (f) = \int\limits_{\sphere} \delta(f(x)) dx,
\end{equation*}
see \eqref{eq:leray meas form}.

Then, taking the expected value of both sides and changing the order
of the expectation and the limit, we obtain
\begin{equation}
\label{eq:ELf=intE} \E\leray (f) = \int\limits_{\sphere}
\E\delta(f(x)) dx.
\end{equation}

Now, for each {\em fixed} $x\in\sphere$, the random variable
$v=f(x)$ is a linear combination of Gaussian random variables, and
therefore, Gaussian itself. Its mean is zero and its variance is $1$
by \eqref{eq:E(f(x)^2)=1}. Writing the Gaussian probability density
function explicitly, we have
\begin{equation*}
\E \delta(f(x)) = \E\delta(v) = \int\limits_{-\infty}^{\infty}
\delta(a) \frac{e^{-\frac{1}{2}a^2}}{\sqrt{2\pi}}da =
\frac{1}{\sqrt{2\pi}}.
\end{equation*}

To finish the proof of this proposition we integrate
the last equality on the sphere and substitute it into
\eqref{eq:ELf=intE}.

\end{proof}

\subsection{Choice of orthonormal bases for $T_{x}(\sphere)$}
\label{sec:orthonorm bas corr mat exp}

For every $x\in\sphere$ we will need to identify
\begin{equation}
\label{eq:Tx(S)=R^m} \phi_{x}: T_{x}(\sphere) \cong \R^{\spheredim},
\end{equation}
so that given a smooth function $f$ on $\sphere$, the function
\begin{equation*}
\nabla f(x) \in\R^{\spheredim},
\end{equation*}
is, under the identification \eqref{eq:Tx(S)=R^m}, {\em almost
everywhere smooth} (i.e, $C^k$, if $f$ is $C^{k+1}$) of argument
$x$.

Since we will be typically interested in the length of the gradient,
we will require the identifications \eqref{eq:Tx(S)=R^m} to be length preserving, namely, isometries.
This is naturally accomplished, given a choice
$$B_{x}=\{e_{1}^{x},\ldots,\, e_{\spheredim}^{x}\}$$ of an
{\em orthonormal} basis of $T_{x}(\sphere)$ for every $x\in\sphere$,
so that for every vector $e_i$, all of its coordinates satisfy the
appropriate smoothness condition. To do so, we consider the sphere
without its southern pole $S$
$$R:=\sphere\setminus \{ S \}.$$ Choosing an {\em arbitrary} orthonormal
basis $B=B_{N}$ corresponding to the northern pole provides such a
basis $B_{x}$ for every $x\in R$ by means of the parallel transport
of $B$ along the unique geodesic linking $N$ and $x$ on $R$. We
choose an arbitrary orthonormal basis $B_{S}$ of the tangent plane
$T_{S}(\sphere)$ of $\sphere$ at the southern pole. It doesn't
affect any of the computations below, and we will neglect it from
this point on.

Let $g(x):\sphere\rightarrow\R$ be any smooth function. We will use
the notation
\begin{equation*}
\frac{\partial}{\partial e_{i}} g (x) = \frac{\partial}{\partial
e_{i}^{x}} g\vert_{x}
\end{equation*}
for the directional derivative of $g(x)$ at $x$ along $e_{i}^{x}$,
i.e.
\begin{equation*}
\frac{\partial}{\partial e_{i}} g (x) = \langle \nabla g (x),
e_{i}^{x} \rangle.
\end{equation*}

In case of ambiguity, i.e. if we deal with a two variable function
\begin{equation*}
h(x,y):\sphere\times\sphere\rightarrow\R,
\end{equation*}
we write $\frac{\partial}{\partial e_{i}^{x}} h$ or
$\frac{\partial}{\partial e_{i}^{y}} h$ for the derivative of $g$ as
a function of $x$ with $y$ constant, or vice versa respectively.
Similarly, we will use the notation $\nabla_{x} g(x,y)\in
T_{x}(\sphere)$ and $\nabla_{y} g(x,y) \in T_{y}(\sphere)$ to denote
the gradient of $g(x,y)$ as a function of $x$ or $y$ respectively.

\begin{remark}
Note that with the choice of the identifications
\eqref{eq:Tx(S)=R^m} as above, we have
\begin{equation}
\label{eq:gradx d = -grady d} \nabla_{x} d(x,y)|_{(x,N)} = -\nabla_{y}
d(x,y)|_{(x,N)},
\end{equation}
which is going to be useful in simplifying the covariance matrix
$\Sigma$ (see section \ref{sec:corr mat, var}).
\end{remark}

\begin{remark}
In fact, for all our purposes, it is sufficient to make the choice
of the orthonormal bases {\em locally}. Such a choice is possible
for any manifold.
\end{remark}

\subsection{The covariance matrix, expectation}
\label{sec:corr mat, exp}

Given a point $x\in\sphere$ we consider the random vector
$(v,w)\in\R\times\R^{\spheredim}$
\begin{equation*}
(v,w) = (f(x),\nabla f(x)),
\end{equation*}
where we use the identification \eqref{eq:Tx(S)=R^m}. It is easy to
see that being a linear transformation of a mean zero Gaussians,
its distribution is a mean zero Gaussian as well.

We claim that the covariance matrix of $(v,w)$ is
\begin{equation}
\label{eq:exp covar mat} \tilde{\Sigma}_{(m+1)\times (m+1)} :=\left(
\begin{matrix}\E f(x)^2 &\E \big[ f(x) \nabla f(x) \big] \\
\E \big[ f(x) \nabla f(x) \big]^{t} &\E
\big[ \nabla f(x) ^ t \nabla f(x) \big] \end{matrix} \right)= \left(\begin{matrix} 1 &0 \\
0 &\frac{\eigval}{\spheredim} I_{\spheredim} \end{matrix}\right).
\end{equation}

First,
\begin{equation}
\label{eq:E(f(x)^2)=1 pr} \E f(x)^2 = u(x,x) = 1,
\end{equation}
by the definition \eqref{eq:u(x,y) def}, \eqref{eq:u(x,y) def ult}
of $u(x,y)$ and \eqref{eq:u(x,y)=1 iff x=pm y}.

Next, we have
\begin{equation*}
\E(f(x)\nabla f(x)) = \frac{1}{2}\nabla \E(f(x)^2) = \nabla 1/2 =
\vec{0}
\end{equation*}
by \eqref{eq:E(f(x)^2)=1}.

Finally, we compute $\E\big[\nabla f(x) ^ t \nabla f(x)\big]$. For $i\ne j$, we
have
\begin{equation*}
\E\bigg[\frac{\partial}{\partial e_i^x} f(x)\frac{\partial}{
\partial e_j^x} f(x) \bigg] = \bigg[\frac{\partial}{\partial e_i^x \partial e_j^y}
 u(x,\,y)\bigg]\bigg|_{x=y} = 0,
\end{equation*}
computing the second partial derivative explicitly in local
coordinates (see section \ref{sec:orthonorm bas corr mat exp} for an
explanation of the partial derivatives notations).

For $i=j$, we have by the rotational symmetry on $\sphere$,
\begin{equation*}
\begin{split}
&\E\bigg(\big(\frac{\partial}{\partial e_i^x} f(x)\big)^2\bigg) =
\frac{1}{\spheredim |\sphere |}\int\limits_{\sphere}\E\bigg(\nabla
f(x) \cdot \nabla f(x)\bigg) dx \\&= \frac{1}{\spheredim |\sphere |}
\E \bigg[\int\limits_{\sphere} \nabla f(x) \cdot \nabla f(x)
dx\bigg] = -\frac{1}{\spheredim |\sphere |} \E
\bigg[\int\limits_{\sphere} f(x) \cdot \triangle f(x) dx\bigg] \\&=
\frac{\eigval}{\spheredim |\sphere |} \E \bigg[\int\limits_{\sphere}
f(x)^2 dx\bigg]= \funcdercorr \cdot
\frac{1}{|\sphere|}\int\limits_{\sphere} \E [f(x)^2] dx =
\funcdercorr,
\end{split}
\end{equation*}
by the divergence theorem and \eqref{eq:E(f(x)^2)=1}. Thus
\begin{equation}
\label{eq:corr grads x} \E(\nabla f(x)^{t} \nabla f(x)) = \E(\nabla
f(y)^{t} \nabla f(y)) = \funcdercorr I_m.
\end{equation}

\subsection{Riemannian volume}

Let $\chi$ be the indicator function of the interval $[-1,1]$. For $\epsilon>0$, we
define the random variable
\begin{equation*}
\length_\epsilon(f):=\frac{1}{2\epsilon} \int_{\sphere}
\chi\bigg(\frac{f(x)}\epsilon\bigg) |\nabla f(x)|dx   \;.
\end{equation*}

\begin{lemma}[Lemma 3.1 from ~\cite{RW}]
\label{lem: formula for Z} Suppose that $f\in \eigspc$ is
nonsingular. Then
$$ \vol(f^{-1}(0))=\lim_{\epsilon \to 0} \length_\epsilon(f) \;.$$
\end{lemma}

Lemma \ref{lem: formula for Z} implies that the expectation of the volume and
its second moments are given by the following.
\begin{corollary}[Corollary 3.4 from ~\cite{RW}]
\label{cor:formulas for moments} The first and second moments of the
volume $\length(f)$ of the nodal set of $f$ are given by
$$\E(\length) = \E(\lim_{\epsilon\to 0} \length_\epsilon),\qquad
\E(\length^2) = \E(\lim_{\epsilon_1,\epsilon_2\to 0} \length_{\epsilon_1}
\length_{\epsilon_2}) \;.
$$
\end{corollary}

\begin{lemma}
\label{lem:Zeps=O(sqrt(E))} For every $f\in\eigspc$ and
$\epsilon>0$, one has
\begin{equation*}
\length_{\epsilon}(f) = O(\sqrt{\eigval}),
\end{equation*}
where the constant involved in the $'O'$ notation depends only on
$\spheredim$.
\end{lemma}

To prove lemma \ref{lem:Zeps=O(sqrt(E))}, we use lemma 3.3 from
~\cite{RW}.

\begin{lemma}[Lemma 3.3 from ~\cite{RW}]\label{lem:Kaclength}
Let $g(t)$ be a trigonometric polynomial on $[0,2\pi]$ of degree at
most $M$. Then for all $\epsilon>0$ we have
$$
\frac 1 {2\epsilon}\int\limits_{\{t: |g(t)|\leq \epsilon
\}}|g'(t)|dt \leq 6M \;.
$$
\end{lemma}

\begin{proof}[Proof of lemma \ref{lem:Zeps=O(sqrt(E))}]

We write $\length_{\epsilon}$ in the multi-dimensional spherical
coordinates (see the proof of lemma \ref{lem:Zeps=O(sqrt(E))}) as
\begin{equation*}
\length_\epsilon(f):=\frac{1}{2\epsilon} \int\limits_{R}
\chi\bigg(\frac{f(\phi_{1},\,\ldots,\,\phi_{\spheredim})}\epsilon\bigg)
\big\|\nabla f(\phi_{1},\,\ldots,\,\phi_{\spheredim})\big\| \cdot
|J| d\phi_{1}\cdot\ldots\cdot d\phi_{\spheredim}.
\end{equation*}
Note that in the spherical coordinates, for $\phi_{k}\ne 0,\pi,2\pi$
\begin{equation*}
\nabla f(\phi_{1},\,\ldots,\, \phi_{\spheredim}) =
\bigg(\frac{1}{\big\|\frac{\partial}{\partial
\phi_{k}}\big\|}\frac{\partial f}{\partial\phi_{k}} \bigg)_{1\le k
\le \spheredim},
\end{equation*}
in the orthonormal basis associated to
$\big\{\frac{\partial}{\partial\phi_{k}}\big\}$. Thus
\begin{equation*}
\|\nabla f \| \cdot |J|\ll  \sum\limits_{k=1}^{\spheredim} \bigg|
\frac{\partial f}{\partial \phi_{k}} \bigg|.
\end{equation*}

Note that for $\phi_{k}$, $k\ne k_{0}$ fixed,
$f(\phi_{1},\,\ldots,\, \phi_{\spheredim})$ is a $1$-variable
trigonometric polynomial in $\phi_{k_{0}}$ of degree $\le
n\ll\sqrt{\eigval}$. Therefore,
\begin{equation*}
\begin{split}
\length_{\epsilon} (f) &\ll \sum\limits_{k_{0}=1}^{\spheredim}
\int\limits_{} \frac{1}{2\epsilon}\bigg[\int\limits_{\{\theta: \:
|f(\phi_{1},\,\ldots,\, \phi_{\spheredim})| < \epsilon \}}
\bigg|\frac{\partial f(\phi_{1},\,\ldots,\,
\phi_{\spheredim})}{\partial\phi_{k_{0}}}\bigg| d\phi_{k_{0}} \bigg]
d\phi_{1} \cdot\ldots \hat{d\phi_{k_{0}}}\ldots \cdot
d\phi_{\spheredim}
\\&\ll \sqrt{\eigval},
\end{split}
\end{equation*}
by lemma \ref{lem:Kaclength}.

\end{proof}

Now we are in a position to prove the main result of this
section, namely proposition \ref{prop:exp len}.

\begin{proof}[Proof of proposition \ref{prop:exp len}]
We saw that
\begin{equation*}
\E\length(f) = \E\lim\limits_{\epsilon\rightarrow 0}
\length_{\epsilon} (f)
\end{equation*}
by corollary \ref{cor:formulas for moments}. Lemma
\ref{lem:Zeps=O(sqrt(E))} and the dominated convergence theorem
allow us to exchange the order of taking expectation and the limit
to obtain
\begin{equation}
\label{eq:Elen=lim Eleneps} \E\length(f) =
\lim\limits_{\epsilon\rightarrow 0} \E \length_{\epsilon}(f).
\end{equation}

By Fubini's theorem,
\begin{equation}
\label{eq:Eleneps=int K1eps} \E \length_{\epsilon}(f) =
\E\bigg[\frac{1}{2\epsilon} \int_{\sphere}
\chi\bigg(\frac{f(x)}\epsilon\bigg) |\nabla f(x)|dx \bigg] =
\int\limits_{\sphere} K_{\epsilon}^{1} (x) dx,
\end{equation}
where $K^{1}_{\epsilon}(x)$ is defined by
\begin{equation}
\label{eq:K1eps def} K^{1}_{\epsilon} (x) := \E \bigg[
\frac{1}{2\epsilon} \chi\bigg(\frac{f(x)}{\epsilon}\bigg) |\nabla
f(x)| \bigg] = \frac{1}{2\epsilon} \int\limits_{\eigspc}
\chi\bigg(\frac{f(x)}\epsilon\bigg) |\nabla f(x)| d\upsilon(f).
\end{equation}

We write $K^{1}_{\epsilon}$ in terms of the random vector $(v,w)$,
introduced in section \ref{sec:corr mat, exp}\footnotemark
\footnotetext{We use this opportunity to note that in ~\cite{RW}, page
7, the boundedness of $\frac{1}{2\epsilon}
\chi(\frac{f(x)}{\epsilon}) | \nabla f(x) |$ is {\em unnecessary},
and, in fact, wrong.} as
\begin{equation*}
K^{1}_{\epsilon} (x) =
\frac{1}{2\epsilon}\int\limits_{\R\times\R^{\spheredim}} \chi\bigg (
\frac{v}{a}\bigg)\|w \| d\mu(v,w),
\end{equation*}
where $d\mu(v,w)$ is the joint probability density function of
$(v,w)$, namely mean zero Gaussian with covariance $\tilde{\Sigma}$
given by \eqref{eq:exp covar mat}. Writing the Gaussian probability
explicitly, we have
\begin{equation*}
\begin{split}
K^{1}_{\epsilon} (x) &=
\frac{1}{2\epsilon}\int\limits_{\R\times\R^{\spheredim}} \chi\bigg (
\frac{v}{\epsilon}\bigg)\|w \| \exp\bigg(-\frac{1}{2}
(v,w)\tilde{\Sigma}^{-1}(v,w)^{t} \bigg)
\frac{dvdw}{(2\pi)^{(\spheredim+1)/2}\sqrt{\det{\tilde{\Sigma}}}}
\\&= \frac{1}{2\epsilon}\int\limits_{-\epsilon}^{\epsilon} \exp\big(-\frac{1}{2}v^2 \big) dv
\int\limits_{\R^{\spheredim}} \|w\| \exp\bigg( -\frac{1}{2}
\frac{\|w\|^2\spheredim}{\eigval}\bigg)\frac{\spheredim^{\spheredim/2}
dw}{(2\pi)^{(\spheredim+1)/2}\eigval^{\spheredim/2}} \\&=
\frac{1}{2\epsilon}\int\limits_{-\epsilon}^{\epsilon}
\exp\big(-\frac{1}{2}v^2 \big) dv \cdot
\frac{\sqrt{\eigval}}{\sqrt{\spheredim}}\int\limits_{\R^{\spheredim}}
\|w'\| \exp\bigg( -\frac{1}{2} \| w'\|^2\bigg)\frac{
dw'}{(2\pi)^{(\spheredim+1)/2}},
\end{split}
\end{equation*}
changing the variables $$w=\sqrt{\frac{\eigval}{\spheredim}}w' .$$

Following \eqref{eq:Eleneps=int K1eps} and \eqref{eq:Elen=lim
Eleneps}, we integrate the last expression and take the limit
$\epsilon\rightarrow 0$ to obtain
\begin{equation*}
\E\length (f) = c_{\spheredim} \sqrt{\eigval},
\end{equation*}
where
\begin{equation}
\label{eq:c exp int} c_{\spheredim} = \frac{|\sphere
|}{\sqrt{\spheredim}(2\pi)^{(\spheredim+1)/2}}\int\limits_{\R^{\spheredim}}
\|w'\| \exp\bigg( -\frac{1}{2} \| w'\|^2\bigg) dw'.
\end{equation}
Finally, substituting $$\int\limits_{\R^{\spheredim}} \|w'\|
\exp\bigg( -\frac{1}{2} \| w'\|^2\bigg) dw' =
\sqrt{2}(2\pi)^{\spheredim/2}\frac{\Gamma\bigg(\frac{\spheredim+1}{2}
\bigg)}{\Gamma\bigg(\frac{\spheredim}{2} \bigg)}$$ (see e.g.
~\cite{RW}, page 7) and \eqref{eq:sphere vol} into the last
expression yields \eqref{eq:c def}.

\end{proof}

\section{An integral formula for the second moment}
\label{sec:int form sec mom}

\subsection{Covariance matrices, second moment}
\label{sec:corr mat, var}

Similarly to the computation of the expected volume, we will
naturally encounter a random vector on
$$\R\times\R\times\R^{\spheredim}\times\R^{\spheredim},$$
defined as
\begin{equation*}
(f(x), f(y), \nabla f(x),\nabla f(y)),
\end{equation*}
for some {\em fixed} $x,y\in\sphere$,
where we again use the identification \eqref{eq:Tx(S)=R^m}. We will use the rotational
symmetry of the sphere to reduce the discussion to the case $y=N$ is the northern pole. Thus
we consider
\begin{equation}
\label{eq:Z rand vec def} Z:=(v_1,v_2,w_1,w_2) = (f(x), f(N), \nabla
f(x),\nabla f(N))
\end{equation}
for some $y\in \sphere$.

It is
obvious that the joint distribution of this vector, is mean zero
Gaussian. It remains, therefore, to compute the covariance matrix.
We need the following notations.

Let $D=D(x)$ be the vector in $\R^{\spheredim}$ defined by
\begin{equation*}
D(x) = \nabla_x u(x,y)|_{(x,N)} \in T_x(\sphere)\cong \R^{\spheredim}.
\end{equation*}
Note that for $x\ne\pm N$, we may use \eqref{eq:u(x,y) def ult} to
obtain $$D(x)= \ultsphpol {'}(d(x,N)) \sin(d(x,N))\nabla_x d(x,\, y)|_{(x,N)}.
$$ It is then clear from \eqref{eq:gradx d = -grady d} that we then have
\begin{equation}
\label{eq:D2=-D1} \nabla_x u(x,y)|_{(x,N)} = -\nabla_{y} u(x,y)|_{(x,N)}.
\end{equation}

Finally, let $$H=H(x)=(h_{ij})$$ be the $\spheredim\times
\spheredim$ matrix defined as
\begin{equation}
\label{eq:pseudo-Hessian of u} H = \nabla_x\nabla_y u(x,y)|_{(x,N)},
\end{equation}
i.e. $H=(h_{jk})$ with entries given by
\begin{equation*}
h_{jk} = \frac{\partial^2}{\partial e_{j}^{x}\partial e_{k}^{y}}
u(x,y)|_{(x,N)}.
\end{equation*}

We will be in particular interested in the conditional distribution
of $$Z_1=(w_1,w_2) = (\nabla f(x),\nabla f(N)),$$ conditioned upon
$f(x)=f(N)=0$.

For the variance computation of the Leray nodal measure, we will
need the distribution of the random vector
\begin{equation*}
\tilde{Z}:=(v_{1},\, v_{2}) = (f(x), f(N)).
\end{equation*}
It is distributed mean zero Gaussian as well.

The covariance matrices of the random vectors above are given in the
following lemma.

\begin{lemma}
\label{eq:Z covar mat} Let $x\in\s$. Then
\begin{enumerate}
\item
\label{it:covar mat ler} The distribution of the random vector
$\tilde{Z}=(v_1,v_2)$ is mean zero Gaussian with covariance matrix given by
\begin{equation}
\label{eq:A blk def} A =  \left(\begin{matrix} 1 &u(x,N) \\ u(x,N)
&1 \end{matrix}\right).
\end{equation}

\item
\label{it:covar mat len} The covariance matrix of the random vector
$Z$ is the $(2\spheredim+2)\times (2\spheredim+2)$ matrix
\begin{equation*}
\Sigma =  \left(\begin{matrix}A & B \\ B^{t} & C
\end{matrix}\right),
\end{equation*}
where $A\in M_{2\times 2}$ is given by \eqref{eq:A blk def}, $B\in
M_{2\times 2\spheredim}$ is given by
\begin{equation}
\label{eq:B blk def} B =  \left(\begin{matrix} \vec{0} &-D(x) \\
D(x) &\vec{0} \end{matrix}\right),
\end{equation} and $C\in M_{2\spheredim\times 2\spheredim}$ is given by
\begin{equation}
\label{eq:C blk def} C =  \left(\begin{matrix} \funcdercorr I_{\spheredim} &H \\
H^{t} &\funcdercorr I_{\spheredim} \end{matrix}\right)
\end{equation}
with the ``pseudo-Hessian" matrix $H=(h_{jk})$ of $u$ given by
\eqref{eq:pseudo-Hessian of u}. The distribution of $Z$ is
nondegenerate for $x\ne\pm N$ (this is equivalent to $\Sigma$ being
invertible).

\item
\label{it:red covar mat} The covariance matrix of the conditional
distribution of $Z_1$, conditioned upon $v_1=v_2=0$ is given by
\begin{equation}
\label{eq:Omega def} \Omega =  \bigg[\left(\begin{matrix}
\funcdercorr I &H \\ H^{t} &\funcdercorr I
\end{matrix}\right) - \frac{1}{1-u^2} \left(\begin{matrix}D^t D &-uD^t D \\ -uD^t D &D^t D \end{matrix}
\right)\bigg].
\end{equation}
We call the matrix $\Omega$ the ``reduced covariance matrix" of
$Z_1$, and one has
\begin{equation}
\label{eq:Jac iden det(Sigma)} \det{\Sigma} = \det{A}\det{\Omega} =
(1-u^2)\det{\Omega}.
\end{equation}
\end{enumerate}
\end{lemma}

\begin{proof}
Part \eqref{it:covar mat ler} of the lemma is evident from the
definition of the two-point function. It is also clear that part
\eqref{it:covar mat len} of the lemma implies part \eqref{it:red covar
mat}, since one computes the covariance matrix $\Omega$ of the
conditional distribution from $\Sigma$ employing $$\Sigma^{-1} =
\left(\begin{matrix}* &* \\ * &\Omega^{-1} \end{matrix} \right).
$$

The nondegeneracy of the distribution of the random vector $Z$ for
$x\ne\pm y$ follows directly from appendix \ref{sec:f (x)(y)gr
f(x)(y) sp}. The matrix $\Sigma$ is then invertible, being the
covariance of a nonsingular joint Gaussian distribution.

It remains, therefore, to prove part \eqref{it:covar mat len} of the
lemma. It is clear that the block $A$ is the same as the covariance
matrix in part \eqref{it:covar mat ler}, i.e. given by \eqref{eq:A blk
def}.

Now by the definition,
\begin{equation*}
B = \left(\begin{matrix} \E(f(x) \nabla f(x)) &\E(f(x)\nabla f(N)) \\
\E(f(N)\nabla f(x)) &\E(f(N)\nabla f(N))\end{matrix}\right),
\end{equation*}
and we have already seen that
\begin{equation*}
\E(f(x)\nabla f(x)) = \vec{0}
\end{equation*}
in section \ref{sec:corr mat, exp} as well as
\begin{equation*}
\E(f(N)\nabla f(N)) = \vec{0}.
\end{equation*}
Also
\begin{equation*}
\E(f(N)\nabla f(x)) = \nabla_x \E(f(x)f(N)) = \nabla_x u(x,y)|_{(x,N)} =
D(x),
\end{equation*}
and similarly
\begin{equation*}
\E(f(x)\nabla f(N)) = -D(x),
\end{equation*}
which finishes the proof of \eqref{eq:B blk def}.

Finally, we compute $C$. By the definition,
\begin{equation*}
C = \left(\begin{matrix} \E(\nabla f(x)^{t} \nabla f(x)) &\E(\nabla f(x)^{t} \nabla f(N)) \\
\E(\nabla f(N)^{t}\nabla f(x)) &\E(\nabla f(N)^{t}\nabla
f(N))\end{matrix}\right).
\end{equation*}

We have already computed that $\E(\nabla f(x)^{t} \nabla f(x))$ and
$\E(\nabla f(N)^{t} \nabla f(N))$ are given by \eqref{eq:corr grads
x}. Finally,
\begin{equation*}
\E(\nabla f(x)^{t} \nabla f(N)) = \nabla_x\nabla_y \E[f(x)f(N)] =
\nabla_x\nabla_y u(x,\, y)|_{(x,N)} = H,
\end{equation*}
and similarly
\begin{equation*}
\E(\nabla f(N)^{t} \nabla f(x)) = H^t.
\end{equation*}
This implies \eqref{eq:C blk def} and finishes the proof of the lemma.

\end{proof}

\subsection{Leray nodal measure}
\label{sec:ler var int form}

\begin{proposition}
\label{prop:ler var int form} The second moment of the Leray nodal
measure is given by
\begin{equation}
\label{eq:ler var int form} \E\leray(f)^2 = \frac{|\sphere
|}{2\pi}\int\limits_{\sphere} \frac{d x}{\sqrt{1-u(x)^2}} ,
\end{equation}
where $u(x)$ is the two-point function given by \eqref{eq:u(x) def}
and \eqref{eq:u(x,y) def ult}.
\end{proposition}

As in the case of expectation, we give a formal derivation of
proposition \ref{prop:ler var int form}, omitting a rigorous
treatment. A rigorous proof is obtained following the lines of the
proof of theorem 5.1 in ~\cite{ORW} (see section 5.3), using lemma
\ref{lem:Sing codim 1 Paxy} in our case. The convergence of the
integral on the RHS of \eqref{eq:ler var int form}, necessary to the
proof, follows from \eqref{eq:var int [-1,1]} and lemma
\ref{lem:1/sqrt(1-u^2) int asymp}.

\begin{proof}[Formal derivation of proposition \ref{prop:ler var int
form}] We write the Leray measure as \eqref{eq:leray meas form}
again, so that
\begin{equation}
\label{eq:ler var dbl atom frm}
\begin{split}
\E \leray(f)^2 &= \E\bigg
[\int\limits_{\sphere}\int\limits_{\sphere} \delta(f(x))\delta(f(y))
dx dy  \ \bigg] = \int\limits_{\sphere\times\sphere} \E\big[
\delta(f(x)) \delta(f(y)) \big] dx dy \\&= |\sphere
|\int\limits_{\sphere} \E\big[ \delta(f(x)) \delta(f(N)) \big] dx,
\end{split}
\end{equation}
by the rotational symmetry of the sphere.
Now, for a {\em fixed} $x\in\sphere$ with $x\ne \pm N$, the random
variables $v_{1}:=f(x)$ and $v_{2} := f(N)$ are multivariate mean zero Gaussian
with covariance matrix $A$ given by \eqref{eq:A blk def}.

Thus, writing the Gaussian measure explicitly, we obtain
\begin{equation*}
\begin{split}
\E\big( \delta(f(x)) \delta(f(y)) &= E \big[\delta(v_1)\delta(v_2)
\big] \\&= \int \limits_{\R^{2}} \delta(a_1) \delta(a_2)
\exp(-\frac{1}{2}a A^{-1}a^{t})\frac{da}{2\pi\sqrt{\det{A}}} \\&=
\frac{1}{2\pi\sqrt{\det{A}}} = \frac{1}{2\pi \sqrt{1-u(x,y)^2}}.
\end{split}
\end{equation*}
Plugging this into \eqref{eq:ler var dbl atom frm} yields
\eqref{eq:ler var int form}.

\end{proof}

\subsection{Riemannian volume}
\begin{proposition}
\label{prop:justif ord chng} The second moment of $\length(f)$ is
given by
\begin{equation}
\label{eq:int form sec mom} \E(\length^2) = |\sphere |\int_{\sphere}
K(x) dx
\end{equation} where
\begin{equation}\label{eq:K(x) def}
K(x) = \frac {1} {\sqrt{1-u^2}}
\int_{\R^{\spheredim}\times\R^\spheredim} \| w_1\| \| w_2\|
\frac{\exp(-\frac {1}{2}  (w_{1},w_{2})\Omega^{-1}
(w_{1},w_{2})^{t})}{\sqrt{\det\Omega}} \frac{dw_1
dw_2}{(2\pi)^{\spheredim+1}},
\end{equation}
where $\Omega=\Omega(x)$ is defined by \eqref{eq:Omega def}.
\end{proposition}

Denote
\begin{equation*}
\label{eq:def K eps(x,y)} K_{\epsilon_1, \epsilon_2} (x,y) :=
\frac{1}{4\epsilon_1\epsilon_2} \int_{\eigspc} \|\nabla f(x) \|
\|\nabla f(y) \| \chi \bigg( \frac{f(x)}{\epsilon_1} \bigg) \chi
\bigg( \frac{f(y)}{\epsilon_2} \bigg) d\upsilon(f) \;.
\end{equation*}

To prove the proposition we will need the following lemma (cf. lemma
5.3 in ~\cite{RW}).

\begin{lemma}\label{lem:bound on K eps(x,y)}
For $(x,\, y)\in\sphere\times\sphere$ with $x\ne y$, one has the
inequality
\begin{equation}
\label{eq:bnd krnl eps} K_{\epsilon_1,\, \epsilon_2} (x,\,
y)\ll_{\spheredim} \frac{\eigval}{\sqrt{1-u(x,y)^2}},
\end{equation}
where the implied constant depends only on the dimension
$\spheredim$.
\end{lemma}

The proof is almost identical to the proof of lemma 5.3 of
~\cite{RW} \footnotemark \footnotetext{Note that in ~\cite{RW}, there is a misprint in the course
of the proof of lemma 5.3.}.

\begin{proof}
Write $f(x) = \langle f,\, U(x) \rangle$, where $U(x)$ is the unit
vector $$U(x) = \sqrt{\funcvalcorrinv} \big(\eta_i (x)\big)_{i} \in
S^{\eigspcdim-1},
$$
where $\big\{ \eta_i (x)\}_{i=1}^{\eigspcdim}$ is the $L_2$
orthonormal basis of $\eigspc$ chosen, and where we identify the
function $f$ with a vector in $\R^{\eigspcdim}$ via \eqref{eq:rand
eigfnc def}. Note that $$\langle U(x),U(y) \rangle = u(x,y)$$ is the
cosine of the angle between $U(x)$ and $U(y)$.

We have
$$\nabla f(x) =  DU \cdot f$$
where the derivative $DU$ is a $\spheredim \times \eigspcdim$
matrix. Equivalently,
$$\big(\nabla f(x)\big)_i =
\bigg\langle f, \bigg(\frac{\partial}{\partial e_i} U(x) \bigg)
\bigg\rangle ,\quad 1\le i\le \spheredim.$$

By the triangle and Cauchy-Schwartz inequalities, $$\|\nabla f (x)
\| \le \sum_{i=1}^{\spheredim} \| f\|\cdot \bigg\|
\bigg(\frac{\partial}{\partial e_i} U(x) \bigg) \bigg\| \ll
\sqrt{\eigval} \| f \|,
$$
due to
\begin{equation*}
\bigg\| \bigg(\frac{\partial}{\partial e_i} U(x) \bigg) \bigg\|^2 =
\E\bigg[ \bigg(\frac{\partial f}{\partial e_{i}}\bigg)^2 \bigg] =
\funcdercorr,
\end{equation*}
by \eqref{eq:corr grads x}.

Therefore
\begin{equation}
\label{eq:mult int Keps} K_{\epsilon_1,\epsilon_2} (x,y) \ll
\frac{\eigval}{4\epsilon_1
\epsilon_2}\int\limits_{\substack{|f(x)| < \epsilon_1 \\
|f(y)|<\epsilon_2}} \|f\|^2 e^{-\| f \| ^2/2} df \;.
\end{equation}

Consider the plane $\pi\subset \R^{\eigspcdim}$ spanned by $U(x)$
and $U(y)$. The domain of the integration is all the vectors
$f\in\R^\eigspcdim$ so that the projection of $f$ on $\pi$ falls
into the parallelogram $P$ defined by the perpendiculars
$l_{x}^{\pm}$ and $l_{y}^{\pm}$ to the endpoints of $\pm U(x)$ and
$\pm U(y)$. Denote the angle $\alpha$ between the sides of $P$,
computed as
$$\cos{\alpha} = \langle U(x), \,U(y) \rangle = u(x,y).$$ We claim
that the area of $P$ is $$ \mbox{area}(P) = 4\epsilon_1 \epsilon_2
\frac{1}{\sqrt{1-u(x,y)^2}} \;.
$$

To see that, we assume, with no loss of generality that
$\epsilon_{2} \cos{\alpha} \le \epsilon_{1}$ (otherwise exchange
between $x$ and $y$) and $\alpha\in (0,\frac{\pi}{2})$. Now if
furthermore,
$$\epsilon_{2} \le \epsilon_{1} \cos{\alpha},$$ then the line $l_{y}^{+}$
does not intersect the interval $[0, \epsilon_{1} U(y)]$, and the
sides of $P$ are easily seen to have lengths
$\frac{2\epsilon_{1}}{\sin{\alpha}}$ and
$\frac{2\epsilon_{2}}{\sin{\alpha}}$, and the angle between the
sides of $P$ is $\alpha$, so that our claim follows. Otherwise
(namely if $\epsilon_{2} > \epsilon_{1} \cos{\alpha}$), a little
trigonometric computation shows that the lengths of the sides of $P$
are again $\frac{2\epsilon_{1}}{\sin{\alpha}}$ and
$\frac{2\epsilon_{2}}{\sin{\alpha}}$ and the angle between the sides
of $P$ is $\alpha$.

Write the multiple integral in \eqref{eq:mult int Keps} as the
iterated integral
\begin{equation}
\label{eq:mult int K rep} \int_{P} \left(\int_{p+\pi^{\perp}} \|
f\|^2 e^{-\| f \|^2/2} df\right) dp \;,
\end{equation}
where the variable $p$ runs over all the points of the
parallelepiped $P$. The inner integral in \eqref{eq:mult int K rep}
is $O(1)$.
% HERE THE CONSTANT DEPENDS ON THE DIMENSION

Indeed, note that for every $f_1\in\pi^{\perp}$,
\begin{equation*}
\begin{split}
\| p+f_1\|^2 e^{-\|p+f_1\| ^2/2}
&= (\|p\|^2+\|f_1\|^2)e^{-(\|p\|^2+\|f_1\| ^2)/2} \\
&\ll (1+\|f_1\|^2) \cdot e^{-\|f_1\| ^2/2}\;,
\end{split}
\end{equation*}
since $\|p\|^2 e^{-\| p \| ^2/2}$ is bounded. Our claim follows from
convergence of the integral $\int_{\R^{\eigspcdim-2}}
(1+\|w\|^2)e^{-\| w\|^2/2}dw $. Therefore
$$
\int_{\substack{|f(x)| < \epsilon_1 \\
|f(y)|<\epsilon_2}} \|f\|^2 e^{-\| f \| ^2/2} df \ll area(P) \ll
\epsilon_1\epsilon_2 \frac{1}{\sqrt{1-u(x,y)^2}} \;.
$$
Substituting the last estimate into \eqref{eq:mult int Keps} proves
\eqref{eq:bnd krnl eps}.

\end{proof}

We give a formal derivation of proposition \ref{prop:justif ord
chng}. Having lemma \ref{lem:bound on K eps(x,y)} in our hands, a
rigorous proof of proposition \ref{prop:justif ord chng} is
identical to the proof of proposition 5.2 of ~\cite{RW} and we omit
it here. In the course of the proof one shows that $$
K(x)=\lim_{\epsilon_1,\epsilon_2\to
0}K_{\epsilon_1,\epsilon_2}(x,N)\;. $$ Therefore, taking the limit
$\epsilon_1,\epsilon_2\to 0$ in \eqref{eq:bnd krnl eps}, we obtain
\begin{corollary}\label{cor:ker bnd ler ker}
If $u(x)^2 \ne 1$ then
\begin{equation*}
K(x) \ll \frac{\eigval}{\sqrt{1-u(x)^2}}\;.
\end{equation*}
\end{corollary}

\begin{proof}[Formal derivation of proposition \ref{prop:justif ord
chng}]

Corollary \ref{cor:formulas for moments} allows us to write an
expression for the second moment formally as
\begin{equation*}
\E\length(f)^2 = \E \bigg[ \int\limits_{\sphere\times\sphere}
\delta(f(x)) \| \nabla f(x)\| \delta(f(y)) \|\nabla f(y) \| dx dy
\bigg],
\end{equation*}
and changing the order of taking the integration, we obtain
\begin{equation}
\begin{split}
\label{eq:sec mom len intexp} \E\length(f)^2 &=
\int\limits_{\sphere\times\sphere} \E\bigg[ \delta(f(x))\cdot \|
\nabla f(x)\| \cdot \delta(f(y))\cdot \|\nabla f(y) \|\bigg] dx dy
\\&= |\sphere|\int\limits_{\sphere} \E\bigg[ \delta(f(N))\cdot \| \nabla
f(N)\| \cdot \delta(f(x))\cdot \|\nabla f(x) \|\bigg] dx,
\end{split}
\end{equation}
by the rotational symmetry of the sphere. In fact, the integrand $$\E\bigg[ \delta(f(x))\cdot \|
\nabla f(x)\| \cdot \delta(f(y))\cdot \|\nabla f(y) \|\bigg]$$ depends on $d(x,y)$ only (this is the
isotropic property of the random ensemble $\eigspc$).

Now for a fixed $x\in\sphere$ with $x\ne\pm
N$, the joint distribution of the random vector $Z$ defined as in
\eqref{eq:Z rand vec def} is Gaussian with mean zero and covariance
$\Sigma = \Sigma(x)$ as in lemma \ref{eq:Z covar mat}. Thus we may
write
\begin{equation*}
\begin{split} &\E\bigg[ \delta(f(x)) \cdot \| \nabla f(x)\|\cdot \delta(f(N)) \cdot\|\nabla
f(N) \|\bigg] \\&= \int\limits_{\R^{2}\times\R^{2\spheredim}}
\delta(v_{1}) \cdot \|w_{1}\| \cdot \delta(v_{2})\cdot \| w_{2} \|
\exp(-\frac{1}{2}(v,w)\Sigma^{-1}(v,w)^{t})\frac{dv
dw}{(2\pi)^{\spheredim+1} \sqrt{1-u^2}\sqrt{\det{\Omega}}},
\end{split}
\end{equation*}
substituting the explicit expression for the Gaussian measure and
using \eqref{eq:Jac iden det(Sigma)} (recall that
$\Omega=\Omega(x)$ is defined by \eqref{eq:Omega def}).

Substituting $v_1=v_2=0$, we have
\begin{equation*}
\begin{split}
&\E\bigg[ \delta(f(x)) \cdot\| \nabla f(x)\|\cdot \delta(f(N))
\cdot\|\nabla f(N) \|\bigg] \\&= \int\limits_{\R^{2\spheredim}}
\|w_{1}\| \| w_{2} \| \exp(-\frac{1}{2}w\Sigma^{-1}w^{t})\frac{
dw}{(2\pi)^{\spheredim+1} \sqrt{1-u^2}\sqrt{\det{\Omega}}}.
\end{split}
\end{equation*}
To obtain the statement of the proposition, we
integrate the last expression over $\sphere$ and
plug it into \eqref{eq:sec mom len intexp}.

\end{proof}

\section{Asymototics of the variance}
\label{sec:asymp var} In this section we prove theorems \ref{thm:var
ler meas} and \ref{thm:var length}.

\subsection{Leray nodal measure}
Here we use the ultraspherical or Gegenbauer polynomials (see
appendix \ref{sec:ultrasph pol} for details).

\begin{proof}[Concluding the proof of theorem \ref{thm:var ler
meas}]

Using proposition
\ref{prop:ler var int form}, \eqref{eq:u(x) def} and proposition \ref{prop:exp ler meas}, we obtain
\begin{equation}
\label{eq:var int [-1,1]}
\begin{split}
\var(\length) = \frac{|\sphere|}{2\pi}\int\limits_{\sphere}
\frac{dx}{\sqrt{1-u(x)^2}} - \frac{|\sphere|^2}{2\pi} =
\frac{|\sphere|}{2\pi}\int\limits_{-1}^{1} \bigg(
\frac{1}{\sqrt{1-\ultsphpol(t)^2}} - 1\bigg)d\mu(t),
\end{split}
\end{equation}
where $\mu=\mu_{\spheredim}$ is the measure on $I:=[-1,1]$ defined
by
\begin{equation} \label{eq:mu def} d\mu(t) =
\frac{2\pi^{\spheredim/2}}{\Gamma(\frac{\spheredim}{2})} \cdot
(1-t^2)^{\frac{\spheredim-2}{2}} dt.
\end{equation}

It is easy to check that $\mu = g_{*}\nu$, where
$g:\sphere\rightarrow I$ is the function
\begin{equation}
\label{eq:g sphr [-1,1] def} g(x):= \cos{d(x,N)},
\end{equation}
and $d$ is the spherical distance (recall that $\nu$ is the uniform
measure on $\sphere$).

Lemma \ref{lem:1/sqrt(1-u^2) int asymp} together with \eqref{eq:var
int [-1,1]} conclude the proof of the theorem once noting
\eqref{eq:eigspcdim asymp}.

\end{proof}

\begin{lemma}
\label{lem:1/sqrt(1-u^2) int asymp} One has the following
asymptotics
\begin{equation*}
\int\limits_{-1}^{1}\bigg[\frac{1}{\sqrt{1-\ultsphpol(t) ^2}} -
1\bigg]d\mu(t) = 2^{m-2} \pi^{m/2} \Gamma(\frac{m}{2})
\frac{1}{n^{\spheredim-1}} + O(\epsilon(\spheredim;n)),
\end{equation*}
where $\epsilon(\spheredim;n)$ is given by
\begin{equation}
\label{eq:eps def} \epsilon(\spheredim;n) := \begin{cases}
\frac{\log{n}}{n^2},\; &\spheredim=2 \\n^{-\spheredim},
&\spheredim\ge 3\end{cases},
\end{equation}
and $\mu$ is the measure defined by \eqref{eq:mu def}.
\end{lemma}

To prove lemma \ref{lem:1/sqrt(1-u^2) int asymp}, we will divide the domain of
the integral (i.e. the interval $I:=[-1,1]$) into two
subintervals: $B := [-1+\frac{c_{0}}{n^2}, 1-\frac{c_{0}}{n^2}]$ with
$c_{0}$ constant, and $B^{c}:= I \setminus B$. We will show
that the main contribution to the integral in \eqref{eq:var int
[-1,1]} comes from $B$, {\em bounding} the contribution of $B^{c}$
to that integral.

We will reuse this partition while proving theorem \ref{thm:var
length} (see section \ref{sec:var len bnd proof}). This justifies
devoting a separate section (namely, section \ref{sec:sing set}) to
the treatment of $B^{c}$. In analogy to the situation of ~\cite{ORW}
(cf. section 6.1) and ~\cite{RW} (cf. section 6.2), we will call $B$
and $B^{c}$ the {\em nonsingular} and the {\em singular} intervals
respectively. The proof of lemma \ref{lem:1/sqrt(1-u^2) int asymp}
will be finally given in section \ref{sec:proof of 1/sqrt(1-u^2)
asymp}.

The singular and nonsingular intervals, as well as some of their
properties will be given in section \ref{sec:sing set}. The proof of
lemma \ref{lem:1/sqrt(1-u^2) int asymp} will be finally given in
section \ref{sec:proof of 1/sqrt(1-u^2) asymp}.

\subsection{The singular interval}
\label{sec:sing set}

In the course of the proofs of theorems \ref{thm:var ler meas} and
\ref{thm:var length}, we are going to deal with the function
$$h(t)=\frac{1}{\sqrt{1-(\ultsphpol(t))^2}}$$ defined on $[-1,1]$. We wish to expand it
into the Taylor polynomial of $f(s)=\frac{1}{\sqrt{1-s^2}}$ around
$s=0$ as
\begin{equation}
\label{eq:1/sqrt(1-u^2) exp nonsing}
\frac{1}{\sqrt{1-(\ultsphpol(t))^2}} = 1+\frac{(\ultsphpol(t))^2}{2}
+ O((\ultsphpol(t))^4).
\end{equation}

To be able to justify the expansion above, we will have to bound
$\ultsphpol(t)$ away from $\pm 1$, as in corollary \ref{cor:|Q|<eps0}.
This corollary provides us with a subinterval $B\subseteq [-1,1]$
(which will be referred as the {\em nonsingular} interval) of large
measure $\mu$, such that $\ultsphpol(t)$ is bounded away from $\pm 1$ for
all $t\in B$. Giving a special treatment to its complement (referred
as the {\em singular} interval, even though it is in fact a union of
two disjoint intervals), we will show that its contribution is
negligible (see sections \ref{sec:sing set contr ler} and
\ref{sec:sing set contr len}). We give a rigorous treatment below.

Let $I$ be the interval $I=[-1,1]$. Choose any $0<\epsilon_0 < 1$
and the constant $c_0>0$ guaranteed by corollary \ref{cor:|Q|<eps0},
corresponding to $\epsilon_0$, assuming that $n$ is large enough in
the sense of corollary \ref{cor:|Q|<eps0}. We fix $\epsilon_0$ and
$c_0$ throughout the rest of the paper and define the nonsingular
interval
\begin{equation*}
B=B_{n} := \big[-1+\frac{c_0}{n^2},1-\frac{c_0}{n^2}\big].
\end{equation*}
Corollary \ref{cor:|Q|<eps0} implies that the expansion
\eqref{eq:1/sqrt(1-u^2) exp nonsing} holds on $B$ with the
constant involved in the $'O'$-notation dependent only on
$\epsilon_0$.

By an explicit computation, it is clear that
\begin{equation}
\label{eq:bnd nonsing set} \mu(B^{c}) \ll n^{- \spheredim},
\end{equation}
where $\mu$ is the measure on $I$ defined by \eqref{eq:mu def}.

Recall that $\mu$ is the measure on $[-1,1]$ induced from the
uniform measure $\nu$ on $\sphere$ by $g:\sphere\rightarrow [-1,1]$
defined by \eqref{eq:g sphr [-1,1] def}. We also define the {\em
spherical} nonsingular set
\begin{equation*}
SB := g^{-1} (B),
\end{equation*}
and the {\em spherical singular set} $$SB^c := \sphere\setminus
SB.$$ Since, as it was mentioned earlier, $\mu=g_{*}\nu$, it is
evident that
\begin{equation}
\label{eq:meas SBc bnd} \nu(SB^c) = \mu(B^c) = O(n^{-\spheredim}).
\end{equation}

The set $SB$ is analogous to the {\em nonsingular} set in the sense
of ~\cite{ORW} (cf. section 6.1) and ~\cite{RW} (cf. section 6.2).
The structure of $SB$ on the sphere (i.e., its projection $B$ into
$[-1,1]$ by $g$) is by far simpler than that of the singular set on
the torus, due to the lack of problems of arithmetic nature.

\subsection{Proof of lemma \ref{lem:1/sqrt(1-u^2) int asymp}}

\label{sec:proof of 1/sqrt(1-u^2) asymp}

\begin{proof}
We write
\begin{equation*}
\int\limits_{-1}^{1}\frac{d\mu(t)}{\sqrt{1-(\ultsphpol(t)) ^2}} dt =
\int\limits_{B} + \int\limits_{B^c}.
\end{equation*}
This, together with lemmas \ref{lem:bnd int 1/sqrt(1-u^2) sing} and
\ref{lem:contr nonsing int 1/sqrt(1-u^2)} imply the result.

\end{proof}

\subsubsection{The contribution of the singular interval $B^{c}$}
\label{sec:sing set contr ler}

\begin{lemma}
\label{lem:bnd int 1/sqrt(1-u^2) sing} One has
\begin{equation}
\label{eq:bnd int 1/sqrt(1-u^2) sing}
\int\limits_{B^c}\frac{d\mu(t)}{\sqrt{1-({\ultsphpol}
 (t))^2}}  \ll n^{-\spheredim}.
\end{equation}
\end{lemma}

\begin{proof}
We will bound the contribution of the integral on $$B^{c}\cap [0,1]
= [1-\frac{c_0}{n^2},1],$$ the rest being similar. Furthermore, we
may assume by symmetry, that $\ultsphpol(t) \ge 0$ so that
$$\frac{1}{\sqrt{1-\ultsphpol(t)^2}} \ll \frac{1}{\sqrt{1-\ultsphpol(t)}} .$$
In what follows we will, consistently with appendix \ref{sec:ultrasph
pol}, adapt the notation $$\alpha:=\frac{\spheredim-2}{2}.$$

Writing $t=\cos{\psi}$, we have $\phi\in [0,\frac{c_1}{n}] $ for
some constant $c_1>0$. Substituting into Hilb's generalized
asymptotic formula (see lemma \ref{lem:Hilb asymp gen}), we have
\begin{equation*}
Q_{n}^{\spheredim} (\cos{\psi}) =
C\cdot\sqrt{\frac{\psi}{\sin{\psi}}}
\frac{J_{\alpha}(n\psi)}{(\sin{\psi})^{\alpha}}+O(\psi^2),
\end{equation*}
for some constant $C=C_{n}^{\spheredim}$, using the normalization
defined by \eqref{eq:Q def}. Taking the limit $\phi\rightarrow 0$,
the value of the constant $C$ is easily seen to be
\begin{equation}
\label{eq:C def} C = \bigg[\lim\limits_{\phi\rightarrow
0}\frac{J_{\alpha}(n\phi)}{\phi^{\alpha}}\bigg]^{-1}=n^{-\alpha}
\tilde{C},
\end{equation}
where
\begin{equation}
\label{eq:tildeC def} \tilde{C} = \tilde{C}^{\spheredim} :=
\bigg[\lim\limits_{\phi\rightarrow
0}\frac{J_{\alpha}(\phi)}{\phi^{\alpha}}\bigg]^{-1}\ne 0,
\end{equation}
since $\ultsphpol(1)\ne 0$ (one can obtain an explicit expression
for this constant using the expansion of the Bessel function into
power series, see e.g. ~\cite{OL}, page 57).

Thus, the contribution of the singular interval to the integral, is,
for $n$ large enough
\begin{equation}
\begin{split}
\label{eq:1/sqrt(1-Q^2) [0,1/n] hilb subs}
&\int\limits_{1-\frac{c_{0}}{n^2}}^{1}\ll\int\limits_{0}^{c_1/n}\frac{(\sin{\phi})^{\spheredim-1}}{\sqrt{1-\ultsphpol(\cos{\phi})}}
d\phi \ll \int\limits_{0}^{c_1/n}
\frac{\phi^{\spheredim-1}}{\sqrt{1-C\cdot\sqrt{\frac{\phi}{\sin{\phi}}}
\frac{J_{\alpha}(n\phi)}{(\sin{\phi})^{\alpha}}+O(\phi^2)}} d\phi
\\&= n^{-\spheredim}\int\limits_{0}^{c_1}
\frac{\psi^{\spheredim-1}}{\sqrt{1-C\cdot (1+O(\frac{\psi}{n})^2)
\frac{J_{\alpha}(\psi)}{(\frac{\psi}{n})^{\alpha}}+O((\frac{\psi}{n})^2)}}
d\psi \\&= n^{-\spheredim}\int\limits_{0}^{c_1}
\frac{\psi^{\spheredim-1}}{\sqrt{1-\tilde{C}
\frac{J_{\alpha}(\psi)}{\psi^{\alpha}}+O((\frac{\psi}{n})^2)}}
d\psi,
\end{split}
\end{equation}
by \eqref{eq:C def}.

We claim that
\begin{equation}
\label{eq:1-CJ>>phi^2} 1-\tilde{C}
\frac{J_{\alpha}(\psi)}{\psi^{\alpha}} \gg_{c_{1}} \psi^{2}.
\end{equation}

Having \eqref{eq:1-CJ>>phi^2} proved would imply that
\begin{equation*}
\int_{1-\frac{c_{0}}{n^2}}^{1} \ll n^{-\spheredim} \int_{0}^{c_{1}}
\psi^{\spheredim-2}d\psi \ll n^{-\spheredim},
\end{equation*}
which is the statement of the lemma.

To see \eqref{eq:1-CJ>>phi^2}, it is sufficient to show that
\begin{equation*}
\lim\limits_{\psi\rightarrow 0} \frac{1-\tilde{C}
\frac{J_{\alpha}(\psi)}{\psi^{\alpha}}}{\psi^2} > 0
\end{equation*}
and
\begin{equation}
\label{eq:|C*J/sin|<1} \bigg|\tilde{C}
\frac{J_{\alpha}(\psi)}{\psi^{\alpha}} \bigg| < 1
\end{equation}
for every $\psi\in (0,c_1]$. However the former inequality follows
from the Bessel function expansion into power series around $\psi=0$
(see ~\cite{OL}, page 57, (9.09))
\begin{equation*}
1-\tilde{C} \frac{J_{\alpha}(\psi)}{\psi^{\alpha}} =
a_0\psi^2+O(\psi^4) ,
\end{equation*}
for some constant $a_0 > 0$, so that the limit is positive.

To see \eqref{eq:|C*J/sin|<1}, we note that in
the course of establishing \eqref{eq:1/sqrt(1-Q^2) [0,1/n] hilb
subs}, we showed
\begin{equation*}
\ultsphpol (\cos{\frac{\psi}{n}}) = \tilde{C}
\frac{J_{\alpha}(\psi)}{\psi^{\alpha}}
+O\bigg(\big(\frac{\psi}{n}\big)^2\bigg).
\end{equation*}
Therefore, if \eqref{eq:|C*J/sin|<1} is not satisfied, taking $n$
large enough would contradict $|\ultsphpol (t)| \le 1$.

\end{proof}

\subsubsection{The contribution of the nonsingular interval $B$}

\begin{lemma}
\label{lem:contr nonsing int 1/sqrt(1-u^2)}
\begin{equation}
\label{eq:contr nonsing int 1/sqrt(1-u^2)}
\int\limits_{B}\bigg[\frac{1}{\sqrt{1-(\ultsphpol(t))^2}} - 1\bigg]
d\mu(t) = 2^{m-2} \pi^{m/2}
\Gamma(\frac{m}{2})\cdot\frac{1}{n^{\spheredim-1}}+O(\epsilon(\spheredim;
n)),
\end{equation}
where $\epsilon(\spheredim;n)$ is given by \eqref{eq:eps def}.
\end{lemma}

\begin{proof}
On $B$ we may write
\begin{equation*}
\frac{1}{\sqrt{1-(\ultsphpol(t))^2}} = 1 +
\frac{(\ultsphpol(t))^2}{2} + O\big(\ultsphpol(t)^4\big)
\end{equation*}
(see section \ref{sec:sing set}). Integrating, we obtain
\begin{equation*}
\begin{split}
&\int\limits_{B}\bigg[\frac{1}{\sqrt{1-(\ultsphpol(t))^2}} - 1
\bigg] d\mu(t) = \frac{1}{2}\int\limits_{B} (\ultsphpol(t))^2
d\mu(t) + O\bigg(\int\limits_{B} (\ultsphpol(t))^4 d\mu(t)\bigg)
\\&= O(\mu(B^c)) + (\frac{1}{2}\int\limits_{-1}^{1}
(\ultsphpol(t))^2 d\mu(t) +O(\mu(B^{c}))) +
O\bigg(\int\limits_{-1}^{1} (\ultsphpol(t))^4 d\mu(t)\bigg) \\&=
\frac{1}{2}(2^{m-1} \pi^{m/2}
\Gamma(\frac{m}{2})\frac{1}{n^{\spheredim-1}}+O(n^{-\spheredim}))+O(\epsilon(\spheredim;n))
\\&=
2^{m-2} \pi^{m/2} \Gamma(\frac{m}{2}) \cdot
\frac{1}{n^{\spheredim-1}}+O(\epsilon(\spheredim;n)),
\end{split}
\end{equation*}
as stated, by \eqref{eq:bnd nonsing set} and lemmas \ref{lem:2nd mom
Qn} and \ref{lem:4th mom Qn}.
\end{proof}

\subsection{Riemannian volume}
\label{sec:var len bnd proof}

The goal of this section is to prove theorem \ref{thm:var length}.

\subsubsection{Plan of the proof of theorem \ref{thm:var length}}

We have by proposition \ref{prop:justif ord chng},
\begin{equation}
\label{eq:val int form} \var(\length(f)) = |\sphere|\int_{\sphere}
K(x) dx - c_{\spheredim}\eigval,
\end{equation}
where
\begin{equation*}
K(x) = \frac {1} {\sqrt{1-u^2}}
\int_{\R^{\spheredim}\times\R^\spheredim} \| w_1\| \| w_2\|
\frac{\exp(-\frac {1}{2}  (w_{1},w_{2})\Omega^{-1}
(w_{1},w_{2})^{t})}{\sqrt{\det\Omega}} \frac{dw_1
dw_2}{(2\pi)^{\spheredim+1}},
\end{equation*}
and $c_{\spheredim}$ is a constant given by \eqref{eq:c def}.

As in case of the Leray nodal measure, we divide the integration
range into the nonsingular set $SB$ and its complement $SB^{c}$
(see section \ref{sec:sing set}). We
bound the corresponding contributions to the integral separately
(see lemmas \ref{lem:int kern Bsing} and \ref{lem:int kern
Bcnonsing}). Using corollary \ref{cor:ker bnd ler ker}, it is easy
to relate the contribution of $SB^{c}$ to the last integral in
\eqref{eq:var int [-1,1]}, which we already bounded while treating
the variance of the Leray nodal measure (lemma \ref{lem:bnd int
1/sqrt(1-u^2) sing}).

It then remains to bound the contribution of the integral on
$SB$. Here we may write $\frac{1}{\sqrt{1-u^2}} = 1+ O(u^2)$ and one
may show that, up to an admissible error, we may replace it by $1$.
We will define a new matrix $S$ by
\begin{equation*}
\Omega = \funcdercorr (I-S),
\end{equation*}
and notice that substituting $S=0$ into the integral, the identity
matrix $I$ recovers the square of the expected volume
$(\E\length)^2$. Bounding the variance is then equivalent to
``bounding" the matrix $S$ in some {\em average} sense.

To quantify the last statement we set $\sigma(x)$ to be the spectral
norm of the matrix $S(x)$. We will show that the variance is bounded
by $$\eigval\cdot\bigg(\int\limits_{\sphere}\sigma(x) dx +
O(\frac{1}{\eigspcdim}) \bigg).$$ To bound $\int\sigma(x)$, we use
the trivial inequality $\sigma(x) \le \sqrt{\tr{S^2}}$. We will
prove that $\int \tr(S(x)^2) \ll \frac{1}{\eigspcdim} $, and
together with the Cauchy-Schwartz inequality this implies the
statement of the theorem.

\subsubsection{A bound for the contribution on the singular interval $SB^c$}
\label{sec:sing set contr len}

\begin{lemma}
\label{lem:int kern Bsing} One has
\begin{equation*}
\int\limits_{SB^{c}} K(x) dx \ll \eigval \epsilon(m;n),
\end{equation*}
where $\epsilon(m;n)$ is defined by \eqref{eq:eps def}.
\end{lemma}

\begin{proof}
We use corollary \ref{cor:ker bnd ler ker} to write
\begin{equation*}
\int\limits_{SB^{c}} K(x) dx \ll_{\spheredim} \eigval
\int\limits_{SB^{c}} \frac{dx}{\sqrt{1-u(x)^2}} = \eigval
\int\limits_{B^{c}} \frac{d\mu(t)}{\sqrt{1-\ultsphpol(t)}} \ll
\eigval\epsilon(\spheredim;n),
\end{equation*}
obtaining the last inequality by lemma \ref{lem:bnd int
1/sqrt(1-u^2) sing}.
\end{proof}

\subsubsection{A bound for the contribution on the nonsingular interval $SB$}

\begin{lemma}
\label{lem:int kern Bcnonsing}
\begin{equation*}
\int\limits_{SB} K(x) dx = \frac{1}{|\sphere|}(\E(\length))^2 +
O(\frac{\eigval}{\sqrt{\eigspcdim}}).
\end{equation*}
\end{lemma}

\begin{proof}
Define $\Omega_1=\Omega_1 (x)$ by $\Omega = \funcdercorr \cdot
\Omega_{1}$. The matrix $\Omega_{1}$ is symmetric, and positive for
a set of $x\ne\pm N$, since $\Omega$ is such.
Therefore it has a positive definite square root $P_{1}^2 =
\Omega_{1}$.

Intuitively, $\Omega_{1}$ approximates the identity matrix $I$. To
quantify this intuitive statement, we introduce the matrix
\begin{equation}
\label{eq:S def} S=I-\Omega_{1} = {\funcdercorrinv} \frac{1}{1-u^2}
\left(\begin{matrix} D^{t} D &-uD ^{t} D-(1-u^2)H \\
-uD^{t} D - (1-u^2)H^{t} & D^{t} D \end{matrix} \right),
\end{equation}
and its spectral norm $\sigma=\sigma(x)$, i.e
$$\sigma = \max\limits_{1 \le i \le 2\spheredim} |\alpha_{i}|,$$ where
$\alpha_i$ are the eigenvalues of $S$. Note that, since $\Omega_{1}$
is positive definite, $S \ll I$ in the sense that all its
eigenvalues are in $(-\infty, 1)$.

Changing the coordinates $$w = \sqrt{\funcdercorr}zP_1,
$$ we write the definition \eqref{eq:K(x) def} of $K(x)$ as
\begin{equation}
\label{eq:K chng var P1} K(x) = \frac{\eigval}{\spheredim
\sqrt{1-u^2}} \int\limits_{\R^{2\spheredim}} \| (zP_{1})_{1}
\| \cdot \| (zP_{1})_{2} \| e^{-\frac{1}{2}\|z \|^{2}}
\frac{dz}{(2\pi)^{\spheredim+1}},
\end{equation}
where for $a\in\R^{2\spheredim}$ we write
$(a)_{1}\in\R^{\spheredim}$ and
$(a)_{2}\in\R^{\spheredim}$ to denote either the first or the
last $\spheredim$ coordinates.

We claim that
\begin{equation}
\label{eq:P1=I+O(sig)} P_{1} = I(1+O(\sigma)).
\end{equation}
This follows from that fact that if $S \sim diag(\alpha_{i})$ then
$P_1 \sim diag(\sqrt{1-\alpha_{i}})$ so that
$$P_1-I\sim diag(\sqrt{1-\alpha_{i}}-1) \ll diag(|\alpha_{i}|) \ll
\sigma I,$$ since $\sqrt{1-y} -1 < |y|$ on $(-\infty, 1)$.

Moreover, by the definition of the spherical nonsingular set, on
$SB$, $u(x)$ is bounded away from $1$, so that one may expand
\begin{equation}
\label{eq:1/sqrt(1-u^2) exp nonsing const} \frac{1}{\sqrt{1-u^2}} =
1+O(u^2),
\end{equation}
where the constant involved in the $'O'$ notation is absolute.

Substituting \eqref{eq:P1=I+O(sig)} and \eqref{eq:1/sqrt(1-u^2) exp
nonsing const} into \eqref{eq:K chng var P1}, we obtain
\begin{equation*}
K(x) = \frac{\eigval}{\spheredim (2\pi)^{\spheredim+1}
}\int\limits_{\R^{2\spheredim}} \| (z)_{1} \| \cdot \|
(z)_{2} \| e^{-\frac{1}{2}\|z \|^{2}}
(1+O(u^2))(1+O(\sigma))^2 dz.
\end{equation*}

Continuing, we have
\begin{equation*}
\begin{split}
K(x) &= \frac{\eigval}{\spheredim (2\pi)^{\spheredim+1} } \int
\limits_{\R^{\spheredim}\times \R^{\spheredim}} \|(z)_{1} \| \cdot \|
(z)_{2} \| e^{-\frac{1}{2}(\|z_{1} \|^{2}+\|z_{2} \|^{2})}
dz_{1}dz_{2} (1+O(u^2)+O(\sigma)+O(\sigma^2))
\\&= \bigg( \frac{\sqrt{\eigval}}{\sqrt{\spheredim}
(2\pi)^{\frac{\spheredim+1}{2}}} \int\limits_{\R^{\spheredim}}
\|z'\| e^{-\frac{1}{2}\|z' \|^{2}} dz' \bigg)^2
(1+O(u^2)+O(\sigma)+O(\sigma^2)) \\&=
\frac{1}{|\sphere|^2}(\E\length)^2 (1+O(u^2)+O(\sigma)+O(\sigma^2)),
\end{split}
\end{equation*}
by \eqref{eq:Elen=c sqrt(E)} and \eqref{eq:c exp int}.

Integrating on $SB$, we obtain
\begin{equation*}
\int\limits_{SB} K(x)dx - \frac{1}{|\sphere|}(\E \length)^2 \ll
\eigval\bigg(\frac{1}{\sqrt{\eigspcdim}}+\int\limits_{SB} u^2dx +
O(\nu(SB^c))\bigg),
\end{equation*}
by \eqref{eq:Elen=c sqrt(E)} and lemma \ref{lem:1st sec mom sigma}.

To bound the last expression, we use \eqref{eq:meas SBc bnd}, as
well as, by the definition \eqref{eq:u(x) def} of the two-point
function, we have
\begin{equation*}
\int\limits_{SB} u^2 dx \le \int\limits_{\sphere} (\ultsphpol
(\cos{d(x,N)}))^2 dx = \int\limits_{-1}^{1} (\ultsphpol (t))^2
d\mu(t) \ll \frac{1}{\eigspcdim},
\end{equation*}
by lemma \ref{lem:2nd mom Qn} and \eqref{eq:eigspcdim asymp}. This
concludes the proof of the lemma.

\end{proof}

\begin{lemma}
\label{lem:1st sec mom sigma} For a fixed $\spheredim$, as
$n\rightarrow\infty$, one has
\begin{enumerate}
\item \label{it:int sigma^2 < 1/N}
\begin{equation*}
\int\limits_{SB} \sigma (x) ^2dx \ll \frac{1}{\eigspcdim}.
\end{equation*}

\item \label{it:int sigma < 1/sqrt(N)}
\begin{equation*}
\int\limits_{SB} \sigma (x)dx \ll \frac{1}{\sqrt{\eigspcdim}}.
\end{equation*}

\end{enumerate}
\end{lemma}

\begin{proof}
Part \ref{it:int sigma < 1/sqrt(N)} of the lemma clearly follows
from part \ref{it:int sigma^2 < 1/N} by the Cauchy-Schwartz
inequality. Thus we are only to prove part \ref{it:int sigma^2 <
1/N}.

To prove the statement, we recall that $\sigma$ is by the definition
the spectral norm of $S$, defined by \eqref{eq:S def}. To bound
$\int\sigma^2$, we use the trivial inequality $\sigma\le \tr(S^2)$.

Since, by the definition of the nonsingular set $SB$, the two-point
function $u(x)$ is bounded away from $1$, we may disregard the $1-u^2$
altogether. We define the matrix
\begin{equation*}
S_{1} := \frac{\eigval}{\spheredim} (1-u^2)S = \left(\begin{matrix} D^{t} D &-uD ^{t} D-(1-u^2)H \\
-uD^{t} D - (1-u^2)H^{t} & D^{t} D \end{matrix} \right).
\end{equation*}
We claim that
\begin{equation*}
\label{eq:tr(S1^2)<<n^3} \int\limits_{\sphere} \tr{S_{1}^2} dx =
O(n^{5-\spheredim}).
\end{equation*}
This is sufficient for the statement of the present lemma, since then
\begin{equation*}
\begin{split} \int\limits_{SB} \sigma^2 dx \ll
\frac{1}{\eigval^2}\int\limits_{SB} \tr{S_1^2} dx \le
\frac{1}{\eigval^2}\int\limits_{\sphere} \tr{S_1^2} dx \ll
\frac{1}{n^4}\cdot n^{5-\spheredim} \ll \frac{1}{\eigspcdim}.
\end{split}
\end{equation*}

Now the elements of the matrix $S^2$ are bounded by elements either
of the form
\begin{equation*}
\frac{\partial u}{\partial e_{i_{1}}^{z}} \vert_{(x,N)}\cdot
\frac{\partial u}{\partial e_{i_{2}}^{z}} \vert_{(x,N)} \cdot
\frac{\partial u}{\partial e_{i_{3}}^{z}} \vert_{(x,N)} \cdot
\frac{\partial u}{\partial e_{i_{4}}^{z}} \vert_{(x,N)},
\end{equation*}
the form
\begin{equation*}
\frac{\partial u}{\partial e_{i_{1}}^{z}} \vert_{(x,N)} \cdot
\frac{\partial u}{\partial e_{i_{2}}^{z}} \vert_{(x,N)} \cdot
\frac{\partial ^{2} u}{\partial e_{i_{3}}^{z} \partial
e_{i_{4}}^{z}} \vert_{(x,N)},
\end{equation*}
or the form
\begin{equation*}
\frac{\partial ^{2} u}{\partial e_{i_{1}}^{z} \partial
e_{i_{2}}^{z}} \vert_{(x,N)} \cdot \frac{\partial ^{2} u}{\partial
e_{i_{3}}^{z}
\partial e_{i_{4}}^{z}} \vert_{(x,N)},
\end{equation*}
where in all the expressions above $z$ may be either $x$ or $y$ (see
section \ref{sec:orthonorm bas corr mat exp} for an explanation of
the partial derivatives notations).

Using the Cauchy-Schwartz inequality again and the symmetry with
respect to the variables, it suffices to prove the inequalities
\begin{equation}
\label{eq:1st der int sph bnd}
\int\limits_{\sphere}\bigg(\frac{\partial u}{\partial e^{x}_1}
(x)\bigg)^4 dx \ll n^{5-\spheredim}
\end{equation}
and
\begin{equation}
\label{eq:2nd der int sph bnd}
\int\limits_{\sphere}\bigg(\frac{\partial^2 u}{\partial
e^{x}_{1}\partial e^{x}_{2}} (x)\bigg)^2 dx \ll n^{5-\spheredim}.
\end{equation}

We may compute the partial derivative in \eqref{eq:1st der int sph
bnd} (assuming $x\ne\pm N$) as
\begin{equation*}
\frac{\partial}{\partial e_{1}^{x}} \ultsphpol (\cos{d(x,N)}) =
-\ultsphpol{'}(\cos{d(x,N)}) \sin{d(x,N)} \frac{\partial}{\partial
e_{1}^{x}} d(x,N),
\end{equation*}
so that, since $\frac{\partial}{\partial e_{1}^{x}} d(x,y) $ is
obviously bounded on $\sphere$, it is sufficient to bound
\begin{equation*}
\int\limits_{\sphere} \big(\ultsphpol{'}(\cos{d(x,N)})
\sin{d(x,y)}\big)^4 dx = \int\limits_{-1}^{1}
\big(\ultsphpol{'}(t)\big)^{4} (1-t^2)^2 d\mu(t),
\end{equation*}
and thus \eqref{eq:1st der int sph bnd} follows from lemma
\ref{lem:4th mom der Qn}, recalling the definition \eqref{eq:mu def}
of the measure $\mu$.

As for \eqref{eq:2nd der int sph bnd}, we write the second partial
derivative in the integrand as
\begin{equation*}
\begin{split}
\frac{\partial^2 u}{\partial e^{x}_{1}\partial e^{x}_{2}} (x) &=
\ultsphpol {''}(\cos{d(x,N)}) \sin^2{d(x,N)}
\frac{\partial}{\partial e_{1}^{x}} d(x,N) \frac{\partial}{\partial
e_{2}^{x}}d(x,N) \\&- \ultsphpol{'}(\cos{d(x,N)}) \cdot
\frac{\partial}{\partial e_{2}^{x}} \bigg[ \sin{d(x,N)}
\frac{\partial}{\partial e_{1}^{x}} d(x,N) \bigg],
\end{split}
\end{equation*}
so that, using a similar argumentation, we conclude that the integral in \eqref{eq:2nd der int sph
bnd} is bounded by
\begin{equation*}
\begin{split}
&\ll \int\limits_{\sphere} \big( \ultsphpol {''}(\cos{d(x,N)})
\big) ^2 (\sin{d(x,N)})^4 dx+\int\limits_{\sphere} \big( \ultsphpol {'}(\cos{d(x,N)}) \big)
^2 dx  \\&= \int\limits_{-1}^{1}\big(
\ultsphpol {''}(t) \big) ^2 (1-t^2)^2 d\mu(t)+\int\limits_{-1}^{1} \big(
\ultsphpol {'}(t)\big)^2 d\mu(t).
\end{split}
\end{equation*}
Therefore, \eqref{eq:2nd der int sph bnd} follows from lemmas
\ref{lem:2nd mom der Qn} and \ref{lem:2th mom sec der Qn}.

\end{proof}

\subsubsection{Concluding the proof of theorem \ref{thm:var length}}

\begin{proof}[Proof of theorem \ref{thm:var length}]
We write \eqref{eq:val int form} as,
\begin{equation*}
\var(\length) =
\int\limits_{SB^c}K(x)dx+\bigg(\int\limits_{SB}K(x)dx-\E(\length)^2\bigg)
\ll \eigval \epsilon(\spheredim; n) +
\frac{\eigval}{\sqrt{\eigspcdim}} \ll
\frac{\eigval}{\sqrt{\eigspcdim}},
\end{equation*}
by lemmas \ref{lem:int kern Bsing} and \ref{lem:int kern
Bcnonsing}, where we use $$\epsilon(\spheredim;n) \ll
\frac{\eigval}{\sqrt{\eigspcdim}},$$ due to \eqref{eq:eps def} and
\eqref{eq:eigspcdim asymp}.
\end{proof}

\appendix

\section{Legendre and ultraspherical polynomials}
\label{sec:ultrasph pol}

The ultraspherical (or Gegenbauer) polynomials
$P^{\alpha}_{n}(t):[-1,1]\rightarrow\R$ of degree $n$ generalize the
Legendre polynomials $P_{n}(t) = P_{n}^{0} (t)$. We use the
corresponding {\em normalized} polynomials $Q^{m}_{n} (t)$ for an
integral $m\ge 0$, which differ from $P^{\alpha}_{n}$ (for a
suitably chosen $\alpha$), by a constant, defined by $$Q^{m}_{n}(1)=1.$$

\subsection{Definition and basic facts}

The Legendre polynomials $P_{n}$ are the unique polynomials of
degree $n$, orthogonal on $[-1,1]$ (w.r.t. the trivial weight
function), normalized by $P_{n}(t)=1$. More generally, for a real
$$\alpha>-1,$$ we define the ultraspherical polynomials
$P_{n}^{\alpha}(t)$, being, up to a constants, the unique sequence
polynomials of degree $n$, pairwise orthogonal w.r.t. the weight
function on $[-1,1]$ defined by
\begin{equation}
\label{eq:omega def}
\omega(t) = (1-t^2)^{\alpha}.
\end{equation}
It
is defined uniquely by the normalizing condition
\begin{equation}
\label{eq:P(1) def} P_{n}^{\alpha}(1) =
\frac{\Gamma(n+\alpha+1)}{\Gamma(n+1) \cdot \Gamma(\alpha+1)},
\end{equation}
once we know that $t=1$ is not a zero of $P_{n}^{\alpha}$, see
~\cite{SZ}, chapter 3.3. The ultraspherical polynomials is a particular case $\alpha=\beta$ of a more
general class of polynomials, usually referred to as the Jacobi
polynomials $P_{n}^{\alpha,\beta}$ (see e.g.
~\cite{SZ} for more information).

While studying the spherical harmonics on the $m$-dimensional
sphere, we are interested in the ultraspherical polynomials with
$\alpha = \frac{m-2}{2}$, and moreover, we would like to normalize
it by setting its value at $1$ to be $1$. That is, we define
\begin{equation}
\label{eq:Q def} Q_{n}^{m}(t) := \frac{P_{n}^{\alpha}
(t)}{P_{n}^{\alpha} (1)},
\end{equation}
where
\begin{equation}
\label{eq:alpha def} \alpha := \frac{m-2}{2}.
\end{equation}
For example,
\begin{equation*}
Q_{n}^{2}(t) = P_{n}(t) = P_{n}^{0}(t)
\end{equation*}
are the usual Legendre polynomials.

Throughout the section, we fix an integral number $m\ge 2$,
and use the associated value of $\alpha$, defined by \eqref{eq:alpha
def}. It is well known that $Q_{n}^{m}$ is either even or odd, for
the even and odd values of $n$ respectively, and $|Q_{n}^{m}(t)|$
has a maximum at $t=\pm 1$.

The function $v=P_{n}^{\alpha}(t)$ satisfies the differential
equation (~\cite{SZ}, page 60, (4.2.1))
\begin{equation}
\label{eq:diff eq ultrsph} (1-t^2)v'' - mtv'+n(n+m-1)v=0.
\end{equation}
Due to its linear nature, it is also satisfied by $v=Q_{n}^{m}(t)$.
The following recurrence relation (~\cite{SZ}, page 83, (4.7.27))
will prove itself as very useful
\begin{equation} \label{eq:recc ultrsph}
(1-t^2)P_{n}^{\alpha} {'} (t)
+ntP_n^{\alpha}(t)-(n+\alpha)P_{n-1}^{\alpha}(t) = 0.
\end{equation}
Note that this recurrence relation is not satisfied by $Q^{m}(t)$
due to the different normalization constants for $P_{n}$ and
$P_{n-1}$.

\subsection{Some basic results}

Recall the definition \eqref{eq:mu def} and ~\eqref{eq:omega def} of
the measure $\mu=\mu_{m}$ and the weight function $\omega_{m}$
respectively. We note that $$d\mu(t) =
\frac{2\pi^{m/2}}{\Gamma(\frac{m}{2})} \cdot \omega(t) dt,$$ so for
purposes of giving an upper bound only, we may disregard the
difference between $d\mu$ and $\omega dt$.

Concerning the $2$nd and the $4$th moments of the ultraspherical
polynomials, we have the following:
\begin{lemma}
\label{lem:2nd mom Qn} For $m$ {\em fixed}, the second moment of the
normalized ultraspherical polynomials is
\begin{equation*}
\int\limits_{-1}^{1} Q_{n}^{m} (t) ^2 d \mu(t) = 2^{m-1} \pi^{m/2}
\Gamma(\frac{m}{2}) \cdot \frac{1}{n^{m-1}} +O(n^{-m}),
\end{equation*}
as $n\rightarrow\infty$.
\end{lemma}

\begin{proof}
One has (~\cite{SZ}, (4.3.3))
\begin{equation}
\label{eq:2nd mom P} \int\limits_{-1}^{1} P^{\alpha}_{n} (t)^2
\omega(t) dt = \frac{2^{m-1}}{2n+m-1}
\frac{\Gamma(n+\frac{m}{2})^2}{\Gamma(n+1)\Gamma(n+m-1)} ,
\end{equation}
and \eqref{eq:P(1) def} implies that
\begin{equation}
\label{eq:P(1) asymp} P^{m}_{\alpha} (1) =
\frac{\Gamma(n+\frac{m}{2})}{\Gamma(n+1)\Gamma(\frac{m}{2})} \sim c
\cdot n^{\alpha}.
\end{equation}
Thus, using the definition \eqref{eq:Q def} of the normalized
ultraspherical polynomials, we obtain
\begin{equation*}
\begin{split}
\int\limits_{-1}^{1} Q^{m}_{n} (t)^2 d\mu(t) &=
\frac{2^{m-1}}{2n+m-1}
\frac{\Gamma(n+1)\Gamma(\frac{m}{2})^2}{\Gamma(n+m-1)} \cdot
\frac{2\pi^{m/2}}{\Gamma(\frac{m}{2})}
\\&= \frac{2^m \pi^{m/2} \Gamma(\frac{m}{2})}{2n+m-1}
\frac{n!}{(n+m-2)!} \\&= 2^{m-1} \pi^{m/2} \Gamma(\frac{m}{2}) \cdot
\frac{1}{n} \cdot\frac{1}{n^{m-2}}(1+O(\frac{1}{n})) \\&= 2^{m-1}
\pi^{m/2} \Gamma(\frac{m}{2}) \frac{1}{n^{m-1}} + O(n^{-m}),
\end{split}
\end{equation*}
as stated.
\end{proof}

\begin{lemma}[Hilb Asymptotics (formula (8.21.17) on page 197 of Szego)]
\label{lem:Hilb asymp gen}
\begin{equation}
\label{eq:Hilb asymp gen} (\frac{1}{2} \sin{\theta})^{\alpha}
P^{\alpha}_{n}(\cos{\theta}) = N^{-\alpha}
\frac{\Gamma(n+\alpha+1)}{n!}
\bigg(\frac{\theta}{\sin{\theta}}\bigg)^{1/2}J_{\alpha}(N\theta)+\delta(\theta),
\end{equation}
uniformly for $0\le\theta\le\pi/2$, where $N=n+\frac{m-1}{2}$,
$J_{\alpha}$ is the Bessel $J$ function of order $\alpha$ and the
error term is
\begin{equation*}
\begin{split} \delta(\theta) \ll \begin{cases}
\theta^{1/2} O(n^{-3/2}), \: &cn^{-1} < \theta < \pi/2 \\
\theta^{\alpha+2}O(n^{\alpha}), \: &0<\theta < cn^{-1}.
\end{cases}
\end{split}
\end{equation*}
\end{lemma}

\paragraph{Remark:} It is clear, that
\begin{equation}
\label{eq:hilb const sim 1} n^{-\alpha}
\frac{\Gamma(n+\alpha+1)}{n!} = 1+O(\frac{1}{n}),
\end{equation}
so we will usually omit this factor.

\begin{corollary}
\label{cor:|Q|<eps0} For every $\epsilon_{0} > 0$, there exists a
constant $c_{0}>0$ depending on $m$ only, such that if $c\ge
c_0$ and $t\in [0,1-\frac{c}{n^2}]$, where $n$ is large enough so
that the interval above is not empty, one has
$$|Q_{n}^{m}(t)| < \epsilon_{0}.$$
\end{corollary}

\begin{proof}
Let $t=\cos{\theta}$. Then if $0 \le t < 1-\frac{c_0}{n^2}$, $\theta
> C_{0}\cdot\frac{\sqrt{c_0}}{n}$ for some absolute constant $C_0>0$.
Lemma \ref{lem:Hilb asymp gen} implies that one has
\begin{equation*}
|P^{\alpha}_{n} (\cos{\theta})| \le C
\frac{1}{\sin^{\alpha}{\theta}} |J_{\alpha}(N \theta)|
\end{equation*}
for some absolute constant $C>0$. We bound it by
\begin{equation*}
|P^{\alpha}_{n} (\cos{\theta})| \le C_1
\frac{n^{\alpha}}{c_0^{\alpha/2}} |J_{\alpha} (N\theta)|,
\end{equation*}
so that \eqref{eq:P(1) asymp} implies that
\begin{equation*}
|Q^{m}_{n} (\cos{\theta})| \le \frac{C_2}{c_0^{\alpha/2}}
|J_{\alpha} (N\theta)| < \epsilon_{0},
\end{equation*}
provided that we choose $c_{0}$ large enough, since $J_{\alpha}$
is bounded.
\end{proof}

\begin{lemma}
\label{lem:4th mom Qn} The $4$th moment of the ultraspherical
polynomials satisfies
\begin{equation*}
\int\limits_{-1}^{1} Q^{m}_{n}(t)^4  d\mu(t) \ll \epsilon(m;n),
\end{equation*}
where $\epsilon(m;n)$ is defined by \eqref{eq:eps def}.
\end{lemma}

\begin{proof}
We will limit ourselves to the interval $[0,1]$. To prove the
statement there, we invoke the generalized Hilb's asymptotics (lemma \ref{lem:Hilb asymp gen}).

We have, using \eqref{eq:hilb const sim 1}, that
\begin{equation*}
(\sin{\theta})^{m-2} (P^{\alpha}_{n}(\cos{\theta}))^4 \ll
J_{\alpha}^4 (N\theta)\frac{\theta^2}{(\sin{\theta})^{m}}  +
\frac{\delta^4(\theta)}{\sin^{m-2}{\theta}}
\end{equation*}
and claim that
\begin{equation}
\label{eq:4th mom P} \int\limits_{-1}^{1} P^{\alpha}_{n}(t)^4
d\mu(t) \ll \begin{cases} \frac{\log{n}}{n^2},\; m=2\\ n^{m-4},\;
m\ge 3\end{cases}.
\end{equation}

We have
\begin{equation}
\label{eq:int Hilb asymp subs} \int\limits_{-1}^{1}
P^{\alpha}_{n}(t)^4 d\mu(t) \ll \int\limits_{0}^{\pi/2}
J_{\alpha}^4(N\theta) \frac{\theta^2}{(\sin{\theta})^{m-1}}d\theta +
\int\limits_{0}^{\pi/2} \frac{\delta^4(\theta)}{\sin^{m-3}{\theta}}
d\theta
\end{equation}

The contribution of the main term in \eqref{eq:int Hilb asymp subs}
to the integral in \eqref{eq:4th mom P} is
\begin{equation*}
\begin{split} &\ll \int\limits_{0}^{\pi/2} J_{\alpha}^4(N\theta)
\frac{1}{\theta^{m-3}}d\theta = N^{m-4}\int\limits_{0}^{\frac{\pi
N}{2}} \frac{J_{\alpha}^4(\phi)}{\phi ^{m-3}} d\phi \\&=
N^{m-4}\bigg[\int\limits_{0}^{1} \frac{J_{\alpha}^4(\phi)}{\phi
^{m-3}} d\phi+\int\limits_{1}^{\frac{\pi
N}{2}}\frac{d\phi}{\phi^{m-1}}\bigg],
\end{split}
\end{equation*}
using the well known decay $$J_{\alpha}(y)\ll\frac{1}{\sqrt{y}} $$
of the Bessel J functions at infinity.

The first integral involved in the expression above is $O(1)$, since
$J_{\alpha}$ vanishes with multiplicity (at least) $\alpha$ at zero
(it follows, for example, from Hilb's formula). The second one is
bounded by
\begin{equation*}
\ll \begin{cases} \log{n},\; m=2 \\ 1,\; m\ge 3\end{cases}.
\end{equation*}
Therefore the contribution of the main term in \eqref{eq:int Hilb
asymp subs} to the LHS of \eqref{eq:4th mom P} is dominated by the
RHS of \eqref{eq:4th mom P}.

The contribution of the error term in \eqref{eq:int Hilb asymp subs}
is at most
\begin{equation*}
n^{2m-4}\int\limits_{0}^{1/n} \theta^{m+7} d\theta +
n^{-6}\int\limits_{1/n}^{\pi/2} \theta^{5-m}d\theta = O(n^{m-12}).
\end{equation*}

We obtain the statement of the lemma by using \eqref{eq:4th
mom P} and \eqref{eq:Q def} with \eqref{eq:P(1) asymp}.

\end{proof}

\subsection{Moments of the derivatives of the ultraspherical polynomials}
\label{sec:mom der Qn}

\begin{lemma}
\label{lem:2nd mom der Qn}
\begin{equation*}
\int\limits_{-1}^{1} Q_{n}^{m}{'}(t)^2 d\mu(t) \ll
\frac{\log{n}}{n^{m-4}}.
\end{equation*}
\end{lemma}

\begin{proof}
We will bound the integral on $[0,1]$ only, having a similar bound
on $[-1,0]$. By \eqref{eq:P(1) asymp}, the statement of the
lemma is equivalent to
\begin{equation*}
\int\limits_{0}^{1} P_{n}^{\alpha}{'}(t)^2 d\mu(t) \ll n^2\log{n}.
\end{equation*}
We rewrite the last integral using \eqref{eq:recc ultrsph} as
\begin{equation*}
\int\limits_{0}^{1} \frac{\bigg((n+m/2-1)P_{n-1}^{\alpha} (t) -
ntP_{n}^{\alpha}(t)\bigg)^2 }{(1-t^2)^2}  d\mu(t).
\end{equation*}

To give a bound, we partition the range of the integration into $2$
subranges:
\begin{equation}
\label{eq:2 subrang 2nd mom der} \int\limits_{0}^{1} =
\int\limits_{0}^{1-1/n^2} + \int\limits_{1-1/n^2}^{1}.
\end{equation}

To bound the second integral in \eqref{eq:2 subrang 2nd mom der}, we
define
\begin{equation}
\label{eq:f numer def} f(t) := (n+m/2-1)P_{n-1}^{\alpha} (t) -
ntP_{n}^{\alpha}(t) = (1-t^2) P_{n}^{\alpha}{'}(t).
\end{equation}
Computing the derivative $f'(t)$ and using \eqref{eq:diff eq
ultrsph} again we obtain
\begin{equation}
\label{eq:f' comp} f'(t) = (m-2)tP_{n}^{\alpha} {'} (t) - n(n+m-1)
P_{n}^{\alpha} (t).
\end{equation}

We claim that this implies
\begin{equation}
\label{eq:f' = O(...)} f'(t) \ll n^2 P_{n}^{\alpha}(1) \ll
n^{\frac{m+2}{2}},
\end{equation}
the second inequality being a consequence of \eqref{eq:P(1) asymp}. To see the first inequality of
\eqref{eq:f' = O(...)}, we note that it is
sufficient to show that
\begin{equation}
\label{eq:tP O est} tP_{n}^{\alpha} {'} (t) \ll n^{\frac{m+2}{2}},
\end{equation}
by \eqref{eq:f' comp}.
With no loss of generality we may assume that $t=1$ or
$P_{n}^{\alpha} {'}$ has a local extremum, i.e. $P_{n}^{\alpha}
{''}(t)=0$. In both cases the equation \eqref{eq:diff eq ultrsph}
implies $$t P_{n}^{\alpha}{'} (t) =  P_{n}^{\alpha}(t) O(n^2),$$
which implies \eqref{eq:tP O est}.

Now using the linear Taylor approximation of $f(t)$ around $t=1$ with
\eqref{eq:f' = O(...)},
the second integral in \eqref{eq:2 subrang 2nd mom der} is, since
$f(1)=0$,
\begin{equation*}
\ll n ^{m+2}\int\limits_{1-1/n^2}^{1} \frac{(t-1)^2}{(1-t^2)^2}
\cdot (1-t^2)^{\frac{m-2}{2}} dt\ll n ^{m+2} \int\limits_{1-1/n^2}^{1}
(1-t^2)^{\frac{m-2}{2}} dt \ll n ^2.
\end{equation*}

In order to bound the first integral in \eqref{eq:2 subrang 2nd mom
der}, we employ the generalized Hilb's asymptotics \eqref{eq:Hilb
asymp gen}. The integrand is (taking the change of variables
$t=\cos{\theta}$ and \eqref{eq:mu def} into the account),
\begin{equation}
\label{eq:hilb appl 2nd mom der}
\begin{split}
&n^2 \frac{\bigg( (1+O(\frac{1}{n})) P_{n-1}^{\alpha} (\cos{\theta})
- \cos{\theta} P_{n}^{\alpha}(\cos{\theta})\bigg)^2
}{(\sin{\theta})^{3}} \cdot (\sin{\theta})^{2\alpha} \\ &\ll
n^2\cdot\frac{\frac{\sin{\theta}}{\theta}\cdot\bigg(\big(1+O(\frac{1}{n})\big)J_{\alpha}((N-1)\theta)-
\big(1+O(\theta^2)+O(\frac{1}{n})\big)J_{\alpha}(N\theta)\bigg)^2}{(\sin{\theta})^3}
+ n^2\frac{\delta^2(\theta)}{(\sin{\theta})^3} \\ &\ll n^2
\frac{\bigg(J_{\alpha}(N\theta) -
J_{\alpha}((N-1)\theta)\bigg)^2}{\theta^3}+
\frac{1}{\theta^3}+O(n^2\theta)
+n^2\frac{\delta^2(\theta)}{\theta^3},
\end{split}
\end{equation}
and the integration range is essentially
$[\frac{1}{n},\frac{\pi}{2}]$.

The contribution of the last error term in \eqref{eq:hilb appl 2nd
mom der} is
\begin{equation*}
\ll n^2 \cdot n^{-3}\int\limits_{1/n}^{\pi/2}
\frac{\theta}{\theta^{3}}d\theta \ll 1,
\end{equation*}
the other ones being trivially bounded by $O(n^2)$.

The contribution of the main term in \eqref{eq:hilb appl 2nd mom
der} is
\begin{equation*}
\begin{split}
&n^2 \int\limits_{1/n}^{\pi/2}
\frac{\big(J_{\alpha}(N\theta)-J_{\alpha}((N-1)\theta)\big)^2}{\theta^3}
d\theta \ll n^4 \int\limits_{1}^{\frac{n\pi}{2}}
\frac{\bigg(J_{\alpha}(\phi)-J_{\alpha}(\phi(1-\frac{1}{N}))\bigg)^2}{\phi^3}d\phi
\\&\ll  n^4\cdot
\frac{1}{n^2}\int\limits_{1}^{\frac{n\pi}{2}}\frac{\phi^2}{\phi^3}d\phi\ll
n^2\log{n},
\end{split}
\end{equation*}
due to the boundness of the derivative $ J_{\alpha}'(t)$. As it was stated, this
is equivalent to the statement of the lemma.
\end{proof}

\begin{lemma}
\label{lem:4th mom der Qn} One has
\begin{equation*}
\int\limits_{-1}^{1} Q_{n}^{m}{'}(t)^4 (1-t^2)^{2} d\mu(t) \ll
\begin{cases} n^{2} \log{n},\; &m=2 \\ \frac{1}{n^{m-4}},\; &m\ge 3 \end{cases}.
\end{equation*}
\end{lemma}

\begin{proof}
The proof of the lemma is similar to the one of lemma
\ref{lem:2nd mom der Qn}.

We will bound the integral only on $[0,1]$, having a similar bound
on $[-1,0]$. The statement of the lemma is equivalent to
\begin{equation*}
\int\limits_{0}^{1} \frac{\bigg((n+m/2-1)P_{n-1}^{\alpha} (t) -
nxP_{n}^{\alpha}(t)\bigg)^4 }{(1-t^2)^2} d\mu(t) \ll
\begin{cases} n^{2}\log{n}, &m=2 \\ n^{m},\; &m\ge 3 \end{cases},
\end{equation*}
using \eqref{eq:P(1) asymp} and \eqref{eq:recc ultrsph}.

We partition the range of the integration into $2$ subranges:
\begin{equation}
\label{eq:2 subrang 4th mom der} \int\limits_{0}^{1} =
\int\limits_{0}^{1-\frac{1}{n^2}} +
\int\limits_{1-\frac{1}{n^2}}^{1}.
\end{equation}

To bound the second integral in \eqref{eq:2 subrang 4th mom der} we
use the definition \eqref{eq:f numer def} of the function $f(t)$, as
well as the inequality \eqref{eq:f' = O(...)}, as in the course of
proof of lemma \ref{lem:2nd mom der Qn}. Thus the integral is
\begin{equation*}
\ll n^{2(m+2)}\int\limits_{1-1/n^2}^{1} \frac{(1-t)^4}{(1-t)^2}
(1-t)^{\frac{m-2}{2}}dt \ll n^{2(m+2)}n^{-(m+4)} = n^m.
\end{equation*}

To bound the first integral in \eqref{eq:2 subrang 4th mom der}, we
employ the generalized Hilb's asymptotics \eqref{eq:Hilb asymp gen}.
The integrand is (taking into consideration the change of variables
$t=\cos{\theta}$),
\begin{equation}
\label{eq:hilb appl 4th mom der}
\begin{split}
&n^4 \frac{\bigg( (1+O(\frac{1}{n})) P_{n-1}^{\alpha} (\cos{\theta})
- \cos{\theta} P_{n}^{\alpha}(\cos{\theta})\bigg)^4
}{\sin{\theta}^{3}\cdot \sin{\theta}^{2\alpha}} \cdot (\sin{\theta})^{4\alpha} \\
&\ll
n^4\cdot\frac{\frac{\sin{\theta}}{\theta}\cdot\bigg(\big(1+O(\frac{1}{n})\big)J_{\alpha}((N-1)\theta)-
\big(1+O(\theta^2)+O(\frac{1}{n})\big)J_{\alpha}(N\theta)\bigg)^4}{(\sin{\theta})^{m+1}}
+ n^4\frac{\delta^4(\theta)}{(\sin{\theta})^{m+1}} \\ &\ll n^4
\frac{\bigg(J_{\alpha}(N\theta) -
J_{\alpha}((N-1)\theta)\bigg)^4}{\theta^{m+1}}+
\frac{1}{\theta^{m+1}}+O(n^4\frac{1}{\theta^{m-7}})
+n^4\frac{\delta^4(\theta)}{\theta^{m+1}},
\end{split}
\end{equation}
and the integration range is (up to a constant)
$[\frac{1}{n},\frac{\pi}{2}]$.

The contribution of the last error term in \eqref{eq:hilb appl 4th
mom der} is
\begin{equation*}
\ll
\frac{n^4}{n^6}\int\limits_{1/n}^{\pi/2}\frac{\theta^2}{\theta^{m+1}}
d\theta = n^{-2}\int\limits_{1/n}^{\pi/2}
\frac{d\theta}{\theta^{m-1}} \ll \max{(n^{m-4}\log{n},1)},
\end{equation*}
the other ones being trivially bounded by $O(n^m)$.

The contribution of the main term in \eqref{eq:hilb appl 2nd mom
der} is
\begin{equation}
\label{eq:4th mom der hilb p=nt}
\begin{split}
&n^4 \int\limits_{1/n}^{\frac{\pi}{2}}
\frac{\big(J_{\alpha}(N\theta)-J_{\alpha}((N-1)\theta)\big)^4}{\theta^{m+1}}
d\theta \ll n^{m+4} \int\limits_{1}^{\frac{\pi N}{2}}
\frac{\bigg(J_{\alpha}(\phi)-J_{\alpha}(\phi(1-\frac{1}{N}))\bigg)^4}{\phi^{m+1}}d\phi.
\end{split}
\end{equation}

Let $g(\phi)$ be the function
\begin{equation*}
g(t) := J_{\alpha}(\phi)-J_{\alpha}(\phi(1-\frac{1}{N})).
\end{equation*}
Then by the mean value theorem,
\begin{equation*}
g(\phi) = \frac{\phi}{N} J_{\alpha}'(s),
\end{equation*}
where $\phi(1-\frac{1}{N}))<s < \phi$, and using the decay
\begin{equation*}
|J_{\alpha}'(s)| \ll \frac{1}{\sqrt{s}},
\end{equation*}
we obtain
\begin{equation}
\label{eq:g decay} |g(\phi)| \ll \frac{\sqrt{\phi}}{N}.
\end{equation}

Substituting \eqref{eq:g decay} into \eqref{eq:4th mom der hilb
p=nt}, we have that the contribution is
\begin{equation*}
\begin{split}
&\ll  n^{m+4}\cdot \frac{1}{n^4}\int\limits_{1}^{\frac{\pi n}{2}}
\frac{\phi^2}{\phi^{m+1}}d\phi \ll n^{m}\int\limits_{1}^{\frac{\pi
n}{2}}\frac{d\phi}{\phi^{m-1}} \ll \begin{cases}  n^{2}\log{n} ,\;
&m=2 \\ n^{m},\; &m\ge 3\end{cases},
\end{split}
\end{equation*}
which concludes the proof of the lemma.

\end{proof}

\begin{lemma}
\label{lem:2th mom sec der Qn}
\begin{equation*}
\int\limits_{-1}^{1} Q_{n}^{m} {''}(t)^2 (1-t^2)^2 d\mu(t) \ll
\frac{1}{n^{m-5}}
\end{equation*}
\end{lemma}

\begin{proof}
We use the differential equation \eqref{eq:diff eq ultrsph} to write
the integral as
\begin{equation*}
\begin{split}
&\int\limits_{-1}^{1} \big(mt Q_{n}^{m}{'}(t)-n(n+m-1)Q_{n}^{m}
(t)\big)^2 d\mu(t) \\ &\ll \int\limits_{-1}^{1}
\big((Q_{n}^{m}{'}(t))^2 d\mu(t) + n^4\int\limits_{-1}^{1} Q_{n}^{m}
(t)^2 d\mu(t) \ll \frac{1}{n^{m-5}},
\end{split}
\end{equation*}
by lemmas \ref{lem:2nd mom Qn} and \ref{lem:2nd mom der Qn}.

\end{proof}

\section{The singular functions are ``rare"}
\label{sec:sing func rare}

In this section we give the proofs of lemmas \ref{lem:Sing codim 1},
\ref{lem:Sing codim 1 Pax} and \ref{lem:Sing codim 1 Paxy} (see
section \ref{sec:sing func}).

\begin{notation}
\label{not:bigcirc,arc,S2} Here and in appendix \ref{sec:f (x)(y)gr
f(x)(y) sp} we adapt the following notations. Let $x$ and $y$ on the
sphere $\sphere$ such that $x\ne\pm y$.

\begin{enumerate}
\item Denote $\bar{xy}$ the (unique) big circle through $x$ and $y$.

\item The smaller arc of $\bar{xy}$ connecting $x$ to $y$ will be denoted
by $\breve{xy}$.

\item Let $z\in\sphere$ be a point not lying on the plane $\Pi=\Pi(x,y)$
defined by $O$, $x$ and $y$. We denote $\mathcal{S}^2 =
\mathcal{S}^2 (x,y,z)$ the (unique) $2$-dimensional big sphere
containing $O$, $x$, $y$ and $z$, i.e. $$\mathcal{S}^2 :=
\sphere\cap\Pi(x,y,z) .$$ (Note that there is no ambiguity in
notations for $\spheredim=2$).

\end{enumerate}

\end{notation}

We also recall the fact that if $\mathcal{S}^2\subseteq\sphere$ is
any big sphere, then for any two points $x,y\in\mathcal{S}^2$,
$$\bar{xy}_{\mathcal{S}^2} = \bar{xy}_{\sphere}.$$ In
particular, the shortest path between $x$ and $y$ on $\sphere$
passes inside $\mathcal{S}^2$ and $$\nabla_{x} d_{\sphere} (x,y)
=\nabla_{x} d_{\mathcal{S}^2} (x,y) \in T_{x} (\mathcal{S}^2)$$
under the natural embedding $$T_{x} (\mathcal{S}^2) \subseteq T_{x}
(\sphere).$$

The following simple geometric lemma will prove itself as quite
useful.

\begin{lemma}
\label{lem:d(x,xi)=d(x,xi'),d(y,xi)=d(y,xi')} Let $x,y\in \sphere$
such that $x\ne y$ and $\xi\ne \xi'\in\sphere$ such that
$d(x,\xi)=d(x,\xi')$ and $d(y,\xi)=d(y,\xi')$. Denote
$v:=\nabla_{x}d(x,y)$ and $$v_{1}=v_{1}(\xi,\xi') = \nabla_{x}
d(x,\xi)-\nabla_{x} d(x,\xi').$$ Then for all $\xi$ and $\xi'$,
$v\perp v_{1}$, and moreover the vectors $v_{1}$ span $v^{\perp}$ in
$T_{x}(\sphere)$.
\end{lemma}

\begin{proof}
To see the claim of the lemma, we first note that it is obvious for
$\spheredim=2$. For higher dimensions, it follows from the fact that
any $2$-dimensional big sphere is given by $\mathcal{S}^2 =
\sphere\cap\Pi$, where $\Pi$ is a $3$-dimensional linear subspace of
$\R^{m+1}$, i.e. one direction vector orthogonal to the plane
containing $\bar{xz}$.
\end{proof}

Recall that the set $Sing\subseteq \eigspc$ is the set of singular
functions (see definition \ref{def:sing func}).

\begin{proof}[Proof of lemma \ref{lem:Sing codim 1}]
We define the map
\begin{equation*}
\Psi:\eigspc\times \sphere\rightarrow \R\times\R^{\spheredim}
\end{equation*}
by
\begin{equation*}
(f,x)\mapsto (f(x), \nabla f(x)),
\end{equation*}
using the isometry $T_{x}(\sphere) \cong \R^{\spheredim}$ again, so
that $$Sing = \pi_{\eigspc} (\Psi^{-1}(0,\vec{0})). $$ We claim that
$\Psi$ is submersion. Having this claim in our hands would imply
\begin{equation*}
\Psi^{-1} (0,\vec{0}) \le \eigspcdim-1
\end{equation*}
by the submersion theorem. Therefore
\begin{equation*}
\dim(Sing) \le \eigspcdim-1
\end{equation*}
as well.

To see that $\Psi$ is indeed a submersion, we compute its
differential to be
\begin{equation*}
d\Psi = \left( \begin{matrix} \eta_{1}(x) &\eta_{2} (x) &\ldots
&\eta_{\eigspcdim} (x) &* \\ \nabla\eta_1(x)  &\nabla\eta_2(x)
&\ldots &\nabla\eta_{\eigspcdim} (x) &
* \end{matrix}\right),
\end{equation*}
where $\{\eta_{k}\}$ is the orthonormal basis of $\eigspc$, which appears in the
definition \eqref{eq:rand eigfnc def} of $f$.
Denote the matrix $A_{(\spheredim + 1)\times \eigspcdim}$ with
the first $\eigspcdim$ columns of $d\Psi$. We claim that $A$ is of
full rank, i.e. $rk(A)=\spheredim+1$. To see that we compute the
Gram matrix of its rows to be
\begin{equation*}
A\cdot A^{t} = \left(\begin{matrix} 1 &0 \\
0 &\frac{\eigval}{\spheredim} I_{\spheredim} \end{matrix}\right),
\end{equation*}
see section \ref{sec:corr mat, exp}. Since it is clearly invertible,
we conclude that $rk(A)=\spheredim+1$.

\end{proof}

Recall that we defined $\PP^{x}_{b}$ and $\PP^{x,y}_{b}$ in section
\ref{sec:sing func} (see \eqref{eq:PPxa def} and \eqref{eq:PPxya
def}).

\begin{proof}[Proof of lemma \ref{lem:Sing codim 1 Pax}]
Define $B^{x}_{b}\subseteq Sing\cap\PP^{x}_{b} $ to be the set of
function having $\pm x$ as their singular point, that is
\begin{equation*}
B^{x}_{b} = \big\{ f\in\PP_{b}^{x}:\: f(x)=0 ,\, \nabla f(x) = 0
\big\} \cup \big\{ f\in\eigspc:\: f(-x)=0 ,\, \nabla f(-x) = 0
\big\}.
\end{equation*}
It is obvious that $B^{x}_{b}$ is nonempty only if $b=0$. Since
$f(-y)=\pm f(y)$ for every $y\in\sphere$, $f\in B^{x}_{b}$ implies
that $f$ is singular at $x$. The set $B^{x}_{b}$ is of
codimension $\spheredim$, since the covariance matrix \eqref{eq:exp covar mat}
is invertible, so that the Gaussian distribution of $\nabla f(x)$ conditioned
upon $f(x)=0$, is nonsingular.

Next, we define $$\bar{B}^{x}_{b}:=
Sing\cap\PP^{x}_{b}\setminus B^{x}_{b}.$$ To prove the statement of the lemma,
we need to prove that $\bar{B}^{x}_{b}$ is of codimension $1$ in
$Sing\cap\PP_{b}^{x}$. To do so, we follow closely the proof of
lemma \ref{lem:Sing codim 1}. This time we define
\begin{equation*}
\Psi_{x}:\eigspc\times \sphere\setminus \{\pm x \} \rightarrow
\R^2\times\R^{\spheredim}
\end{equation*}
by
\begin{equation*}
(f,y)\mapsto (f(x),f(y), \nabla f(y)),
\end{equation*}
satisfying
\begin{equation*}
\bar{B}^{x}_{b} = \pi_{\eigspc} (\Psi^{-1}(b,0,\vec{0})).
\end{equation*}

Using a similar dimensional approach, it is sufficient to prove that
$\Psi_{x}$ is a submersion. The differential of $\Psi_{x}$ is
\begin{equation*}
d\Psi_{x} = \left( \begin{matrix} \eta_{1}(x) &\eta_{2} (x) &\ldots
&\eta_{\eigspcdim} (x) &* \\ \eta_{1}(y) &\eta_{2} (y) &\ldots
&\eta_{\eigspcdim} (y) &*\\ \nabla\eta_1(y)  &\nabla\eta_2(y)
&\ldots &\nabla\eta_{\eigspcdim} (y) &
* \end{matrix}\right).
\end{equation*}
Assume by contradiction that the vectors $(\eta_{k}(x),\eta_{k}
(y),\, \nabla\eta_{k}(y))$ satisfy a nontrivial linear functional. Since
$\eta_{k}$ span the whole space $\eigspc$, that functional is
satisfied by
$$\big(\ultsphpol(\cos{d(x,\xi)}),\ultsphpol(\cos{d(y,\xi)}),\nabla
\ultsphpol(\cos{d(y,\xi)})\big),$$ for every $\xi\in\sphere$.
The surjectivity of $d\Psi_{x}$ then follows from lemma \ref{lem:dPsix
surj eq}.

Finally, we note that
\begin{equation*}
Sing\cap\PP^{x}_{b} = B^{x}_{b}\cup \bar{B}^{x}_{b},
\end{equation*}
which concludes the proof of this lemma.
\end{proof}

\begin{lemma}
\label{lem:dPsix surj eq} For every $x\in\sphere$, $y\ne \pm x$, the
only solutions in $\alpha,\beta\in\R$, $C\in\R^{\spheredim}$ for
\begin{equation}
\label{eq:bas rel singpxa} \alpha Q_{n}^{m}(\cos{d(x,\xi)})+\beta
Q_{n}^{m}(\cos{d(y,\xi)})- Q_{n}^{m}{'}(\cos{d(y,\xi)})
\sin{d(y,\xi)} \langle C,\, \nabla_{y} d(y,\xi)\rangle = 0
\end{equation}
are $\alpha=\beta=0$, $C=\vec{0}$.
\end{lemma}

\begin{proof}
It is obvious that either $\alpha=0$ or $\beta=0$, imply that
$\alpha=\beta=0$, $C=\vec{0}$. Therefore we may assume that
$\alpha=-1$, $\beta\ne 0$.

Substituting $\xi$ in \eqref{eq:bas rel singpxa} and, in addition, any $\xi'\ne
\xi$ not lying on $\bar{xy}$ with $d(\xi',x)=d(\xi,x)$ and
$d(\xi',y)=d(\xi,y)$, we obtain that $C$ is collinear to any
$v_1=v_{1}(\xi,\xi')\in T_{y}(\sphere)$ of the form $$v_{1} =
\nabla_{y}d(y,\xi)-\nabla_{y}d(y,\xi').$$ Lemma
\ref{lem:d(x,xi)=d(x,xi'),d(y,xi)=d(y,xi')} implies that $C$ is
collinear to $v:=\nabla _{y} d(x,y)$.

We restrict ourselves to any two-dimensional big sphere
$\mathcal{S}^2 \subseteq \sphere$ containing $x$ and $y$. Knowing
that $C\parallel v$, for $\xi\in\mathcal{S}^2$ on the big circle
perpendicular to $\nabla_{y} d(x,y)$, \eqref{eq:bas rel singpxa} is
\begin{equation*}
Q_{n}^{m} (bt) = \frac{1}{\beta} Q_{n}^{m} (t),
\end{equation*}
by the spherical cosine theorem, where we denote $t:=\cos{d(y,\xi)}$
and $b:=\cos{d(x,y)}$. It is clear that it is only possible if
$b=1$, that is $x=\pm y$, which is a contradiction.

\end{proof}

\begin{proof}[Proof of lemma \ref{lem:Sing codim 1 Paxy}]
To prove the statement of the lemma, we partition the set
$Sing\cap\PP^{x,y}_{b}$ into
\begin{equation*}
Sing\cap\PP^{x,y}_{b} = B^{x}_{b} \cup B^{x,y}_{b}\cup
\bar{B}^{x,y}_{b},
\end{equation*}
defining appropriately each of the sets above and proving the
statements regarding each of them separately.

First, similarly to the proof of lemma \ref{lem:Sing codim 1 Pax},
we treat the set $$B^{x}_{b}\subseteq Sing\cap\PP^{x,y}_{b} $$ of
function having $\pm x$ as their singular point, that is
\begin{equation*}
B^{x}_{b} = \big\{ f\in\PP_{b}^{x,y}:\: f(x)=0 ,\, \nabla f(x) = 0
\big\} \cup \big\{ f\in\eigspc:\: f(-x)=0 ,\, \nabla f(-x) = 0
\big\}.
\end{equation*}
It is easy to see (exactly as in case of lemma \ref{lem:Sing codim 1
Pax}) that $B^{x}_{b}$ has codimension $\ge 1$ in $\PP^{x,y}_{b}$.

Next, we treat the case when the function $f$ has a singular point
on $D\subseteq \sphere$, a distinguished codimension $1$ set of
points on the sphere we are about to define. Let $A_1 := \bar{xy}$
be the big circle linking $x$ to $y$ and $A_2\subseteq \sphere$
be the set of all the points $z$ such that the spherical angle $\angle xzy =
\frac{\pi}{2}$ is right angle. Define $$D=D_{x,y}=A_1\cup A_2
\setminus \{ \pm x \}.$$ It is clear that $D$ is a codimension one
set on the sphere satisfying $\pm x\notin
D$, $\pm y \in D$.

Define $B^{x,y}_{b}\subseteq Sing\cap\PP^{x,y}_{b} $ to be the set
of singular functions having a $D$-point as their singular point, that is
\begin{equation*}
B=B^{x,y}_{b} = \big\{ f\in\PP_{b}^{x,y}:\: \exists z\in D:f(z)=0
,\, \nabla f(z) = 0 \big\}.
\end{equation*}

We claim that $B$ has codimension at least $1$ in $\PP_{b}^{x,y}$.
To see that we define the map
\begin{equation*}
\tilde{\Psi}_{b_1}^{x,y}:\eigspc\times
D\rightarrow\R^{2}\times\R^{\spheredim}
\end{equation*}
by
\begin{equation*}
(f,z)\mapsto (f(x),f(z),\nabla f (z)).
\end{equation*}
It is clear that $$B \subseteq \pi_{\eigspc}
(({\tilde{\Psi}^{x,y}_{b_1}}) ^{-1} (b_{1},0,\vec{0})).$$ Moreover,
$\tilde{\Psi}^{x,y}_{b_1}$ is a submersion (see the proof of lemma
\ref{lem:Sing codim 1 Pax}).

Therefore, $({\tilde{\Psi}^{x,y}_{b_1} }){^{-1}} (b_{1},0,\vec{0})$ is
of codimension $\spheredim+2$ in $\eigspc\times D$, i.e. of
dimension $\eigspcdim-3$, so that $B$ is of codimension $\ge 1$ in
$\PP_{a}^{x,y}$.

Finally, we treat the ``generic" case. We define the set
\begin{equation*}
\bar{B}=\bar{B}^{x,y}_{b} := Sing\cap\PP^{x,y}_{b}\setminus
(B^{x}_{b} \cup B^{x,y}_{b})
\end{equation*}
of functions in $Sing\cap\PP^{x,y}_{b}$ having the set of their
singular points outside of $\{\pm x\}\cup D$ (i.e. having
at least one singular point there).

We define
\begin{equation*}
{\Psi}_{b}^{x,y}:\eigspc\times \sphere \setminus (D\cup \{\pm
x\})\rightarrow\R^{3}\times\R^{\spheredim}
\end{equation*}
by
\begin{equation*}
(f,z)\mapsto (f(x),f(y),f(z),\,\nabla f (z)).
\end{equation*}
It is obvious that
\begin{equation*}
\bar{B} = \pi_{\eigspc} (({\Psi}^{x,y}_{b_1}) ^{-1}
(b_{1},b_{2},0,\vec{0})).
\end{equation*}

As before, to prove that $\bar{B}$ is of codimension $1$, it is
sufficient to prove that ${\Psi}_{b}^{x,y}$ is a submersion. To see
that ${\Psi}_{b}^{x,y}$ is a submersion, we compute its differential
to be
\begin{equation*}
d\Psi = \left( \begin{matrix} \eta_{1}(x) &\eta_{2} (x) &\ldots
&\eta_{\eigspcdim} (x) &* \\ \eta_{1}(y) &\eta_{2} (y) &\ldots
&\eta_{\eigspcdim} (y) &* \\ \eta_{1}(z) &\eta_{2} (z) &\ldots
&\eta_{\eigspcdim} (z) &* \\ \nabla\eta_1(z)  &\nabla\eta_2(z)
&\ldots &\nabla\eta_{\eigspcdim} (z) &
* \end{matrix}\right).
\end{equation*}
Its surjectivity follows from lemma \ref{lem:dPsixy surj eq}.

This concludes the proof of this lemma.

\end{proof}

\begin{lemma}
\label{lem:dPsixy surj eq} Let $x$, $y$ and $z$ be points on the
sphere $\sphere$. Suppose that $x\ne y$, $z\notin \bar{xy}$, and
$\angle xzy \ne \frac{\pi}{2}$. Then the only solution to
\begin{equation}
\label{eq:bas lin dep 3 pnt} \alpha f(x)+\beta f(y)+\gamma
f(z)+\langle C,\, \nabla_{z} f(z) \rangle = 0
\end{equation}
for every $f\in\eigspc$ is $\alpha=\beta=\gamma = 0 $, $C=\vec{0}$.
\end{lemma}

\begin{proof}
Substituting
\begin{equation*}
f(z) = \ultsphpol(\cos{d(z,\, \xi)})
\end{equation*}
with $\xi \in\sphere$, \eqref{eq:bas lin dep 3 pnt} is
\begin{equation}
\label{eq:bas lin dep 3 pnt subs Pn}
\begin{split}
&\alpha \ultsphpol (\cos{d(x,\, \xi)})+\beta \ultsphpol (\cos{d(y,\,
\xi)})+\ultsphpol (\cos{d(z,\, \xi)})\\&-\ultsphpol{'} (\cos{d(z,\,
\xi)})\sin{d(z,\, \xi)}\langle C,\, \nabla_{z} d(z,\xi) \rangle = 0.
\end{split}
\end{equation}
for $\xi \ne \pm z$.

Comparing the equality \eqref{eq:bas lin dep 3 pnt subs Pn} for
$\xi$ not lying on $\bar{yz}$ and any $\xi'$ with
$d(x,\xi)=d(x,\eta')$ and $d(z,\xi)=d(z,\xi')$, we obtain
\begin{equation}
\label{eq:bas relat subs harm symm pair}
\begin{split}
&\ultsphpol {'}(\cos{d(z,\xi)})\sin(d(z,\xi))\langle C,\nabla_{z}
d(z,\xi)- \nabla_{z} d(z,\xi')\rangle \\&= \beta \bigg[\ultsphpol
(\cos{d(y,\xi')}) - \ultsphpol (\cos{d(y,\xi)}) \bigg].
\end{split}
\end{equation}

Let $$\xi''\ne\xi'''\in\sphere$$ with $\xi''\ne \xi$ be the unique pair of points with
\begin{equation*}
d(z,\xi'')=d(z,\xi''')=d(z,\xi)
\end{equation*}
and
\begin{equation*}
\nabla_{z}d(z,\xi'') - \nabla_{z}d(z,\xi''')=\nabla_{z}d(z,\xi) -
\nabla_{z}d(z,\xi').
\end{equation*}
In particular, we have
\begin{equation*}
d(x,\xi'')=d(x,\xi''').
\end{equation*}

Substituting $\xi''$ and $\xi'''$ into \eqref{eq:bas relat subs harm
symm pair}, as we may do, yields
\begin{equation}
\label{eq:bas relat subs harm dual symm pair}
\begin{split}
&\ultsphpol {'}(\cos{d(z,\xi)})\sin(d(z,\xi))\langle C,\nabla_{z}
d(z,\xi'')- \nabla_{z} d(z,\xi''')\rangle \\&= \beta
\bigg[\ultsphpol(\cos{d(y,\xi''')}) - \ultsphpol (\cos{d(y,\xi'')})
\bigg],
\end{split}
\end{equation}
and comparing \eqref{eq:bas relat subs harm symm pair} to
\eqref{eq:bas relat subs harm dual symm pair} we see that either
$\beta = 0$ or
\begin{equation}
\label{eq:4 symm pnts sat} \ultsphpol (\cos{d(y,\xi)}) - \ultsphpol
(\cos{d(y,\xi')}) - \ultsphpol (\cos{d(y,\xi'')}) + \ultsphpol
(\cos{d(y,\xi''')})=0.
\end{equation}

Suppose by contradiction that the latter holds. We restrict
ourselves to any big two-dimensional sphere
$\mathcal{S}^2(x,y,z)\subseteq\sphere$ (recall notation
\ref{not:bigcirc,arc,S2}). Let $d>0$ be a small number and
$\phi\ne\phi'\in \mathcal{S}^2$ be the (unique) points which satisfy
\begin{equation*}
d(z,\phi) = d(z,\phi') = d
\end{equation*}
and $\bar{\phi\phi'}\perp \bar{xz}$. We may approach to $\phi$ by
$\mathcal{S}^2$-points $\xi$ and $\xi''$ and to $\phi'$ by $\xi'$
and $\xi'''$ of the form as above with the additional requirement
\begin{equation*}
d(\xi,\bar{xz})=d(\xi',\bar{xz})=d(\xi'',\bar{xz})=d(\xi''',\bar{xz})=d.
\end{equation*}

Dividing \eqref{eq:4 symm pnts sat} by $d(\xi,\xi'') =
d(\xi',\xi''')$, and taking the limit as $\xi\rightarrow\phi$, we
obtain
\begin{equation}
\label{eq:4 symm pnts sat lim}
\ultsphpol (\cos{d(y,\phi)}) \sin{d(y,\phi)}
\frac{\partial}{\partial e^{\phi}} d(y,\phi) = \ultsphpol
(\cos{d(y,\phi')}) \sin{d(y,\phi')} \frac{\partial}{\partial
e^{\phi'}} d(y,\phi '),
\end{equation}
where $e^{\phi}$ and $e^{\phi'}$ are the unit tangent vectors in the
directions $\breve{\phi\xi''}$ and $\breve{\phi'\xi'''}$ respectively.

Denote $d_0:=d(y,z) $, $d_1 := d(y,\phi) $ and $d_2:=d(y,\phi')$.
Let $\delta$ be the angle $\delta = \angle xzy$. We have $\delta\ne
0,\frac{\pi}{2}$ by the assumptions of the lemma. We compute
\begin{equation*}
c_1=c_1(d):=\cos{d_1} = \cos d_0\cos d +
\cos{\delta}\sin{d_0}\sin{d}
\end{equation*}
and
\begin{equation*}
c_2=c_2(d):=\cos{d_2} = \cos d_0\cos d
-\cos{\delta}\sin{d_0}\sin{d},
\end{equation*}
by the spherical cosine theorem. It is obvious that the LHS
of \eqref{eq:4 symm pnts sat lim} is an
analytic function of $c_1$, and the RHS is the same function
evaluated at $c_2=g(c_{1})$ for some analytic function $g$. The function
$g$ is defined on an neighbourhood of $\cos{d_{0}}$ satisfying $g(\cos{d_{0}})=\cos{d_{0}}$.
Therefore, lemma \ref{lem:P(g(x))=P(x)=>g'=+-1}
implies that
\begin{equation*}
g'(\cos{d_0}) = \pm 1.
\end{equation*}

On the other hand, computing the derivative explicitly, we have
\begin{equation*}
g'(\cos{d_0}) =
\frac{-\cos{d_0}\sin{d_0}-\cos{\delta}\cos{d_0}\sin{d_0}}{-\cos{d_0}\sin{d_0}+\cos{\delta}\cos{d_0}\sin{d_0}}
= \frac{1+\cos{\delta}}{1-\cos{\delta}},
\end{equation*}
which, clearly, under the assumptions of the lemma, cannot be equal
to $\pm 1$, and therefore we obtain the necessary contradiction. This
proves that $\beta=0$. By the symmetry, we have $\alpha=0$ as well.

Thus \eqref{eq:bas relat subs harm symm pair} implies that
$$C\perp v_{1} (\xi):=\nabla_{z} d(z,\xi)-\nabla_{z} d(z,\xi'),$$ for
every $\xi,\xi'$ of the form above. However, for every $\xi$, the
vectors $$v_1(\xi,\xi') \in T_{z}(\sphere)$$ are all orthogonal to
$v:=\nabla_{z} d(x,z)$, the vector in the direction of $\bar{xz}$,
and moreover, they span the orthogonal complement $v^{\perp}$, by
lemma \ref{lem:d(x,xi)=d(x,xi'),d(y,xi)=d(y,xi')}.

Therefore $C$ must be collinear to $v$. Similarly we may argue that
$C$ is collinear to $v':=\nabla_{z} d(z,y)$. However, $v$ and $v'$
are not collinear by the assumptions of the present lemma, so that $C=0$. Knowing that,
$\gamma=0$ is easy to obtain.

\end{proof}

\begin{lemma}
\label{lem:P(g(x))=P(x)=>g'=+-1}

Let $f(t)$ an analytic, not identically vanishing function, and $g(t)$ a differentiable function defined
on an neighbourhood $I$ of $t_0\in I$ such that $g(t_0)=t_0$.
Suppose that we have on $I$
\begin{equation*}
f(g(t))=f(t).
\end{equation*}
Then $g'(t_0)=\pm 1$.
\end{lemma}

\begin{proof}
We have by the chain rule,
\begin{equation*}
f'(g(t))g'(t) = f'(t).
\end{equation*}
Therefore, if $f'(t_0)\ne 0$ then $g'(t_0)=1$ and we are done.
Otherwise, we continue differentiating to obtain
\begin{equation*}
f''(g(t))g'^2(t) +f'(g(t))g''(t)= f''(t)
\end{equation*}
so that if $f''(t_0)\ne 0$, we have $g'^2(t_0)=1$ and we are done
again. Otherwise we continue differentiating until we encounter the
first derivative $f^{(k)}(t_0) \ne 0$ implying $g'^{k}(t) = 1$. Such
a number $k$ exists, since $f$ is {\em analytic}.

\end{proof}

\section{Non degeneracy of point value and gradient distribution}
\label{sec:f (x)(y)gr f(x)(y) sp}

In this section we prove that for $\pm N \ne x \in\sphere$, the
distribution of the random vector $Z$ defined in section
\ref{sec:corr mat, var}, is {\em nonsingular} Gaussian.

\begin{lemma}
Let $x \ne \pm N\in\sphere$ and $V=V_{x}$ be vector space
\begin{equation*}
V = \R^2\times T_{x} (\sphere)\times T_{N}(\sphere).
\end{equation*}
Define the subspace
\begin{equation*}
U = U_{x,\, n} \subseteq V
\end{equation*}
by
\begin{equation*}
U = \{\big(f(x),\,f(N),\, \nabla f(x),\, \nabla f(y) \big)
:\:f\in\eigspc\}.
\end{equation*}
Then one has
$$U = V,$$ provided that $n$ large enough. That is, the distribution of the
random vector
\begin{equation*}
V=\big(f(x),\,f(N),\, \nabla f(x),\, \nabla f(N) \big)
\end{equation*}
is Gaussian nondegenerate and one may identify
\begin{equation*}
U \cong \R^{2\spheredim+2},
\end{equation*}
as in section \ref{sec:orthonorm bas corr mat exp}.

\end{lemma}

\begin{proof}
Let $x\ne \pm N$. We assume by contradiction, that $U$ is a proper
subspace of $V$, i.e. there is a nontrivial functional
$h:V\rightarrow\R$ vanishing on $U$.

We wish to work with coordinates and employ the orthonormal bases for
$T_{x}(\sphere)$ and $T_{N} (\sphere)$ chosen in section \ref{sec:orthonorm bas corr mat exp},
so that under the corresponding identification, one has \eqref{eq:gradx d = -grady d}.

By our assumption, there exist numbers $\alpha,\, \beta\in\R$ and
vectors $C,\, D\in\R^{2}$ so that
\begin{equation}
\label{eq:bas relat} \alpha f(x)+\beta f(N) + \langle C,\,
\nabla f(x) \rangle + \langle D,\, \nabla f(N) \rangle = 0.
\end{equation}
We know that for every $\eta\in\sphere$, the function
\begin{equation*}
f(x):= \ultsphpol(\cos{d(x,\eta)}),
\end{equation*}
is a spherical harmonic lying in $\eigspc$. For this particular
function \eqref{eq:bas relat} is for $\eta\ne \pm x,\pm N$,
\begin{equation}
\label{eq:bas relat subs harm}
\begin{split} &\alpha \ultsphpol(\cos{d(x,\,
\eta)})+\beta \ultsphpol(\cos{d(N,\, \eta)}) \\&- \ultsphpol
{'}(\cos{d(x,\,\eta)})\sin{(d(x,\,\eta))}\cdot\langle C,\,
\nabla_{x} d(x,\eta) \rangle \\&- \ultsphpol
{'}(\cos{d(N,\,\eta)})\sin{(d(N,\,\eta))}\cdot\langle D,\,
\nabla_{N} d(N,\eta) \rangle = 0.
\end{split}
\end{equation}

First choose $\eta\in\sphere$ not lying on $\bar{xN}$ and compare
\eqref{eq:bas relat subs harm} for $\eta$ and any $\eta'\ne\eta$
satisfying $d(x,\,\eta') = d(x,\,\eta)$ and $d(y,\,\eta') =
d(y,\,\eta)$. We obtain
\begin{equation*}
\begin{split}
&\ultsphpol {'} (\cos{d(x,\,\eta)})\sin{(d(x,\,\eta))}\cdot\langle
C,\, \nabla_{x} d(x,\eta) \rangle \\&+ \ultsphpol {'}
(\cos{d(N,\,\eta)})\sin{(d(N,\,\eta))}\cdot\langle D,\, \nabla_{N}
d(N,\eta) \rangle \\&= \ultsphpol {'}
(\cos{d(x,\,\eta)})\sin{(d(x,\,\eta))}\cdot\langle C,\, \nabla_{x}
d(x,\eta') \rangle \\&+ \ultsphpol {'}
(\cos{d(N,\,\eta)})\sin{(d(N,\,\eta))}\cdot\langle D,\, \nabla_{N}
d(N,\eta') \rangle.
\end{split}
\end{equation*}

Equivalently,
\begin{equation}
\label{eq:relat subs sides chng}
\begin{split}
&\ultsphpol {'} (\cos{d(x,\,\eta)})\sin{(d(x,\,\eta))}\cdot\langle
C,\, \nabla_{x} d(x,\eta) -\nabla_{x} d(x,\eta')\rangle \\&=
-\ultsphpol {'}(\cos{d(N,\,\eta)})\sin{(d(N,\,\eta))}\cdot\langle
D,\, \nabla_{N} d(N,\eta) -\nabla_{N} d(N,\eta')\rangle.
\end{split}
\end{equation}

For every $\eta$, the vectors $$v_1(\eta) = \nabla_{x} d(x,\eta) -
d(x,\eta') \in T_{x}(\sphere)$$ are all orthogonal to $v:=\nabla_{x}
d(x,N)$, the vector in the direction of $\bar{xN}$, and moreover,
they span the orthogonal complement $v^{\perp}$ by lemma
\ref{lem:d(x,xi)=d(x,xi'),d(y,xi)=d(y,xi')}. We claim that the
equality \eqref{eq:relat subs sides chng} implies that $$C \perp
sp\{v_1 (\eta)\}$$ and thus $C$ and $v$ are collinear. Similarly,
$D$ and $v':=\nabla_{N} d(x,N)$ are collinear, and since we identify $v$ with
$-v'$, that implies $C=\lambda D$ are collinear.

Suppose otherwise. Let $v_{0} = v_{1} (\eta_{0})$ such that $\langle
C,\, v_{0} \rangle \ne 0$, and consider the two-dimensional sphere
$\mathcal{S}^2\subseteq\sphere$ defined by $\bar{xN}$ and $v_0$. For
$\eta\in \mathcal{S}^2$, one has $$v_{1}(\eta)\parallel v_{0}.$$ We
fix $d=d(N,\eta)$ so that $\cos{d}$ is a zero of $\ultsphpol {'}$.
Then the RHS of \eqref{eq:relat subs sides chng} vanishes and our
assumptions imply that $\cos{d(N,\eta)}$ is a zero of $\ultsphpol
{'}$.

However the function $\cos{d(x,\,\eta)}$ is a continuous nonconstant
function of $\eta$ on the arc
$$A:=\{\eta :\: d(N,\eta)=d \}\subseteq\mathcal{S}^2,$$ and therefore its image contains an
interval, contradicting the finiteness of number of zeros of
$\ultsphpol {'}$. Therefore $$C \parallel v,$$ which proves our
claim, i.e. $C=\lambda D$ for some $\lambda\in \R$.

Substituting the last equality into \eqref{eq:relat subs sides chng}
with $\eta\in\sphere$ such that $$d(x,\eta)=d(N,\eta),$$ implies
$\lambda = -1$, i.e.
\begin{equation}
\label{eq:C=-D} C=-D .
\end{equation}

Now substitute $\eta\rightarrow x$ in \eqref{eq:bas relat subs harm}
to obtain
\begin{equation}
\label{eq:bas relat subs harm eta=x} \alpha +\beta \ultsphpol
(\cos{d}) + \ultsphpol {'}(\cos{d})\sin{(d)}\cdot\langle C,\,
\nabla_{N} d(N,x) \rangle = 0,
\end{equation}
where $d=d(x,N)$. We obtain similarly
\begin{equation}
\label{eq:bas relat subs harm eta=y} \alpha \ultsphpol (\cos{d})
+\beta - \ultsphpol {'} (\cos{d})\sin{(d)}\cdot\langle C,\,
\nabla_{x} d(x,N) \rangle = 0,
\end{equation}
upon substitution $\eta\rightarrow N$. Since in our identification,
we have $\nabla_{x} d(x,N) = - \nabla_{N} d(x,N)$, \eqref{eq:bas
relat subs harm eta=x} together with \eqref{eq:bas relat subs harm
eta=y} imply
\begin{equation}
\label{eq:alpha=beta} \alpha=\beta,
\end{equation}
since $$(\ultsphpol (\cos{d}) \ne 1) \Leftarrow (d\ne 0,\pi)
\Leftrightarrow (x\ne\pm N).$$

We claim that $\alpha=0$ and $C=0$. Assume otherwise. Consider any
two-dimensional sphere $\mathcal{S}^2$ containing $\bar{xN}$, and
the big circle $E\subseteq \mathcal{S}^2$ defined by
\begin{equation*}
E = \{\eta\in\mathcal{S}^2:\: d(x,\,\eta) = d(N,\,\eta) \}.
\end{equation*}
On $E$, \eqref{eq:bas relat subs harm} is, substituting
\eqref{eq:C=-D} and \eqref{eq:alpha=beta}
\begin{equation}
\label{eq:bas relat subs harm on E} 2\alpha \ultsphpol
(\cos{d(x,\eta)}) + \ultsphpol {'} (\cos{d(x,\eta)}) \sin{d(x,\eta)}
\langle C,\, \nabla_{N} d(N,\, \eta) -\nabla_{x} d(x,\, \eta)
\rangle = 0.
\end{equation}
It is clear that the vector $$v=\nabla_{N} d(N,\, \eta) -\nabla_{x}
d(x,\, \eta)$$ is collinear to $\nabla_x d(x,N)$, which, as we have
seen, collinear to $C$. In particular, $\alpha =0$ if and only if
$C=0$ and thus we may assume by contradiction $\alpha\ne 0$ and
$C\ne 0$.

Since $x\ne \pm N$, the point $\eta\in E$ lying on $\breve{xN}$,
satisfies $$d(x,\eta) < \frac{\pi}{2}$$ and the point $\eta' = -\eta
\in E$ satisfies
$$d(x,\eta) > \frac{\pi}{2}.$$ Therefore there exists a point
$\eta_0 \in E$ with $$d(x,\eta_{0}) = \frac{\pi}{2}.$$ Then either
$\ultsphpol(\cos{d(x,\eta_{0})})=0$ or $\ultsphpol {'}
(\cos{d(x,\eta_{0})})=0$, depending on whether $n$ is even or odd.
However, the equality \eqref{eq:bas relat subs harm on E} implies
then
\begin{equation*}
\ultsphpol (\cos{d(x,\eta_{0})})=\ultsphpol {'}
(\cos{d(x,\eta_{0})})=0.
\end{equation*}
This contradicts the fact that $\ultsphpol$ does not have any double
zeros, since then the differential equation \eqref{eq:diff eq
ultrsph} satisfied by $\ultsphpol$ would imply $\ultsphpol \equiv
0$.

\end{proof}

\end{document}